\documentclass[aps,prb,reprint,superscriptaddress,amsmath,amssymb,amsthm,mathrsfs]{revtex4-1}

\usepackage{graphicx}
\usepackage{dcolumn}
\usepackage{bm}
\usepackage[ 
margin=0.75in,
]{geometry}

\usepackage[usenames]{color}

\def\sjba#1{}
\def\bafa#1{}

\DeclareMathAlphabet{\mathpzc}{OT1}{pzc}{m}{it}
\def\bakxy{Ba$_{1-x}$K$_x$(Zn$_{1-y}$Mn$_y$)$_2$As$_2$}
\def\bak{(Ba,K)(Zn,Mn)$_2$As$_2$}
\def\Tc{$T_{\mathrm{C}}$}
\def\musr{$\mu$SR}

\begin{document}
	
	\title{Local atomic and magnetic structure of dilute magnetic semiconductor (Ba,K)(Zn,Mn)$_2$As$_2$}
	
	\author{Benjamin A. Frandsen}
	\affiliation{Department of Physics, Columbia University, New York, NY 10027, USA}

	\author{Zizhou Gong}
	\affiliation{Department of Physics, Columbia University, New York, NY 10027, USA}
	
	\author{Maxwell Terban}
	\affiliation{Department of Applied Physics and Applied Mathematics, Columbia University, New York, NY 10027, USA}

	\author{Soham Banerjee}
	\affiliation{Department of Applied Physics and Applied Mathematics, Columbia University, New York, NY 10027, USA}

	\author{Bijuan Chen}
	\affiliation{ %
		Institute of Physics, Chinese Academy of Sciences, Beijing, China
	} %
	
	\author{Changqing Jin}
	\affiliation{ %
		Institute of Physics, Chinese Academy of Sciences, Beijing, China
	} %

	\author{Mikhail Feygenson}
	\affiliation{J\"{u}lich Center for Neutron Science, Forschungszentrum J\"{u}lich GmbH, D-52425, J\"{u}lich, Germany.}
	
	\author{Yasutomo J. Uemura}
	\affiliation{Department of Physics, Columbia University, New York, NY 10027, USA}
		
	\author{Simon J. L. Billinge}
	\email{sb2896@columbia.edu}
	\affiliation{%
		Condensed Matter Physics and Materials Science Department, Brookhaven National Laboratory, Upton, NY 11973, USA
	}%
	\affiliation{Department of Applied Physics and Applied Mathematics, Columbia University, New York, NY 10027, USA}
	
	\date{\today}
	
	\begin{abstract}
	We have studied the atomic and magnetic structure of the dilute ferromagnetic semiconductor system \bak\ through atomic and magnetic pair distribution function analysis of temperature-dependent x-ray and neutron total scattering data. We detected a change in curvature of the temperature-dependent unit cell volume of the average tetragonal crystallographic structure at a temperature coinciding with the onset of ferromagnetic order. We also observed the existence of a well-defined local orthorhombic structure on a short length scale of $\lesssim 5$~\AA, resulting in a rather asymmetrical local environment of the Mn and As ions. Finally, the magnetic PDF revealed ferromagnetic alignment of Mn spins along the crystallographic $c$-axis, with robust nearest-neighbor ferromagnetic correlations that exist even above the ferromagnetic ordering temperature. We discuss these results in the context of other experiments and theoretical studies on this system.
	
	\end{abstract}
	
	
	\maketitle
	
	\section{Introduction}
	Dilute ferromagnetic semiconductors (DFS's) are materials that exhibit semiconducting charge transport behavior in conjunction with ferromagnetic order induced by a small amount of magnetic impurities~\cite{dietl;rmp14}. Over the last two decades, the unique combination of transport and magnetic characteristics in these materials has made them the subject of significant fundamental research into dilute magnetism, as well as a major component of the modern applied field of ``spintronics,'' in which both charge and spin degrees of freedom are exploited to achieve novel functionalities in devices. Historically, doped III-V semiconductors such as (Ga,Mn)As have received the most attention~\cite{ohno;s98}, but within the past few years, another family of DFS's based on the chemistry and structure of the iron-based superconductors has been discovered, including compounds such as the ``111''-type Li(Zn,Mn)As,\cite{deng;nc11} ``122''-type \bak,~\cite{zhao;nc13} and others.\cite{chen;prb14,deng;prb13,ning;prb14,zhao;jap14,ding;prb13} These materials possess a number of attractive characteristics, including the ability to independently control the charge-carrier and spin concentrations and the availability of bulk-form and single-crystal specimens. In contrast, Mn-doping in (Ga,Mn)As simultaneously introduces spins and charge carriers and leads to severe chemical insolubility, restricting specimens with an appreciable Mn concentration to metastable epitaxial thin films.
	
	Despite a general consensus that the ferromagnetic exchange between localized dilute magnetic moments in DFS systems is mediated by the itinerant holes, the microscopic details of the mechanism for ferromagnetism and the possible routes to attaining a higher \Tc\ are still unclear. A significant obstacle has been the lack of bulk-form specimens necessary for detailed characterization by such workhorse techniques as powder x-ray diffraction, neutron scattering, and NMR. These new DFS materials, available as bulk powder specimens and in some cases as single crystals, therefore provide an outstanding opportunity to gain more detailed experimental information and further advance our theoretical understanding of DFS materials.
	
\begin{figure}
		\includegraphics[width=80mm]{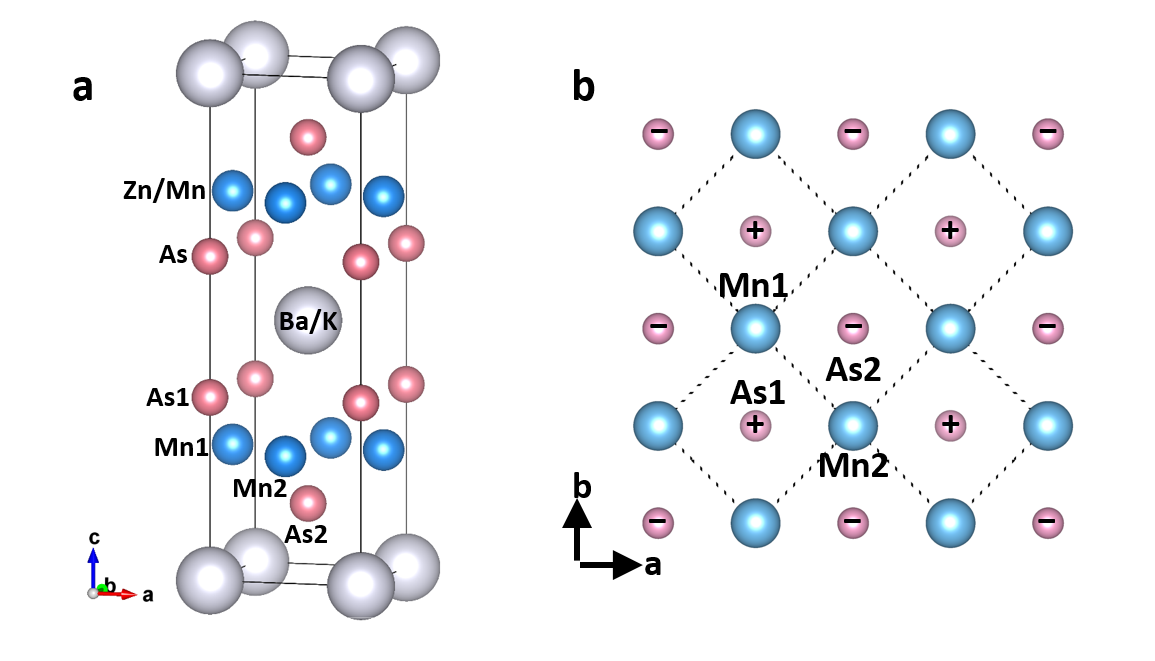}
		\caption{\label{fig:struc} (Color online) Crystal structure of \bakxy. (a) Average crystallographic unit cell determined by x-ray and neutron diffraction experiments. (b) View of the Zn(Mn)/As layer from above. The dashed lines represent the square planar network of nearest neighbor Mn atoms. The $+$ and $-$ signs indicate whether the As atom is located above or below the Mn plane, respectively. The numerical labels assigned to the atoms are explained in the text, and are to be understood in reference to the Zn(Mn)/As plane in the lower half of the unit cell.}
	\end{figure}

	Of the new DFS materials discovered so far, \bakxy\ has the highest \Tc, reaching 230~K for optimal dopant concentrations ($x=0.3$ and $y=0.15$).~\cite{zhao;csb14} This is comparable to the highest \Tc\ of state-of-the-art films of (Ga,Mn)As. The heterovalent substitution of K$^{1+}$ for Ba$^{2+}$ introduces holes, while the isovalent substitution of Mn$^{2+}$ for Zn$^{2+}$ introduces spins. Initial characterization by x-ray and neutron diffraction revealed a phase-pure material possessing the layered tetragonal ThCr$_2$Si$_2$ crystal structure (space group I4/mmm, displayed in Fig.~\ref{fig:struc}), and magnetization and muon spin relaxation (\musr) measurements revealed ferromagnetism developing throughout the full sample volume with a saturated magnetic moment of 1-2~$\mu_{\mathrm{B}}$ per Mn atom~\cite{zhao;nc13}.
	Follow-up studies~\cite{suzuk;prb15a,suzuk;prb15b} using x-ray absorption and photoemission clarified the electronic structure of this system, indicating that metallic transport occurs via doped holes residing primarily in the As 4$p$ valence band and confirming the $S=5/2$, Mn$^{2+}$ nature of the Mn dopants. Two first-principles studies of this system have been performed~\cite{glasb;prb14,yang;sss15}, both suggesting that the hole-mediated ferromagnetic interactions competes with an antiferromagnetic superexchange interaction between nearest-neighbor (NN) Mn ions mediated by As anions.
	
	A key piece of experimental data that has been missing is a thorough, temperature-dependent study of the atomic and magnetic structure of this material. The x-ray diffraction measurements used to initially verify the crystal structure and phase purity were performed only at room temperature, and the subsequent neutron powder diffraction experiment was conducted on a sparse temperature grid that served to verify the I4/mmm average crystal structure at room temperature and below but could offer only limited insight into the structural behavior as the temperature was lowered through \Tc.~\cite{zhao;nc13} Moreover, the low-temperature neutron diffraction patterns were insufficient for determining the magnetic structure, presumably due to the dilute and disordered distribution of Mn ions, leaving open the question of the orientation of the spins.
	
	Here, we present temperature-dependent atomic and magnetic pair distribution function (PDF) measurements of several compositions of \bak\ obtained from synchrotron x-ray and neutron time-of-flight total scattering experiments. This represents the first PDF investigation of local structure in a DFS material. The atomic PDF data reveal a well-defined local orthorhombic structure coexisting with the long-range, average tetragonal structure. At a temperature coinciding with the onset of long-range magnetic order, the temperature dependence of the average tetragonal unit cell volume exhibits an anomaly, reminiscent of spontaneous volume magnetostriction in common ferromagnets like nickel, and the local structure shows a slight change in the orthorhombicity and the Mn-As-Mn angles. The magnetic PDF (mPDF) measurements reveal ferromagnetic alignment of Mn spins along the crystallographic $c$-axis, with significant short-range ferromagnetic correlations between nearest-neighbor Mn spins persisting up to at least 300~K. We discuss these results in the context of existing theories and experimental data on \bak\ and related materials.
	
	\section{Experimental Details}
	Powder samples of \bakxy\ with compositions ($x,y$) = (0,~0), (0,~0.15), (0.2,~0.15), (0.3,~0.15), and (0.4,~0.15) were synthesized using the arc-melting solid-state reaction method, as described previously~\cite{zhao;nc13,zhao;csb14}. The phase purity was verified via x-ray diffraction with a Philips X'pert diffractometer using copper K-edge radiation. For ($x,y$)=(0.3,~0.15), a very small impurity phase of Ba$_2$Mn$_3$As$_2$O$_2$ was found to exist at the level of $\lesssim$ 1 weight percent. No impurity phases were detected for the other compositions. The DC magnetic susceptibility for each sample was measured with a Quantum Design SQUID magnetometer.
	
	Temperature-dependent x-ray total scattering measurements were performed on the samples with ($x,y$)=(0,~0) and (0.3,~0.15) at beamline X17A of the National Synchrotron Light Source (NSLS) at Brookhaven National Laboratory (BNL), with the temperature controlled by a gas-flow cryostream system. Additional x-ray total scattering measurements were performed on the samples with ($x,y$)=(0,~0) and (0.2,~0.15) at the XPD beamline of the National Synchrotron Light Source II (NSLS-II) at a single temperature of 100~K. Neutron time-of-flight total scattering measurements were conducted on the samples with ($x,y$)=(0,~0.15), (0.2,~0.15) and (0.4,~0.15) at the NOMAD instrument~\cite{neufe;nimb12} at the Spallation Neutron Source (SNS) of Oak Ridge National Laboratory (ORNL), with the temperature controlled by an Orange Cryostat. The x-ray data were reduced and transformed to the real-space PDF using the program xPDFsuite~\cite{yang;arxiv15} with a maximum momentum transfer of $Q_{\mathrm{max}}$=25~\AA$^{-1}$. The neutron data were reduced and transformed with $Q_{\mathrm{max}}$=20~\AA$^{-1}$ using the automatic data reduction scripts at the NOMAD beamline. The atomic PDF modeling was carried out using the PDFgui program~\cite{farro;jpcm07} and the SrFit module within the DiffPy library of diffraction analysis software~\cite{juhas;aca15}. The mPDF modeling was performed with the diffpy.mpdf package available as part of the DiffPy library. Estimates of the standard uncertainty of the refined parameters of the least-squares fits were obtained according to the protocol in Appendix B of Ref.~\onlinecite{yang;jac14} for the case of experimental data for which the full variance-covariance matrix is not propagated through the data reduction process, as is the \textit{de facto} situation for PDF experiments.
	
	\section{Results}
	
\begin{figure}
		\includegraphics[width=60mm]{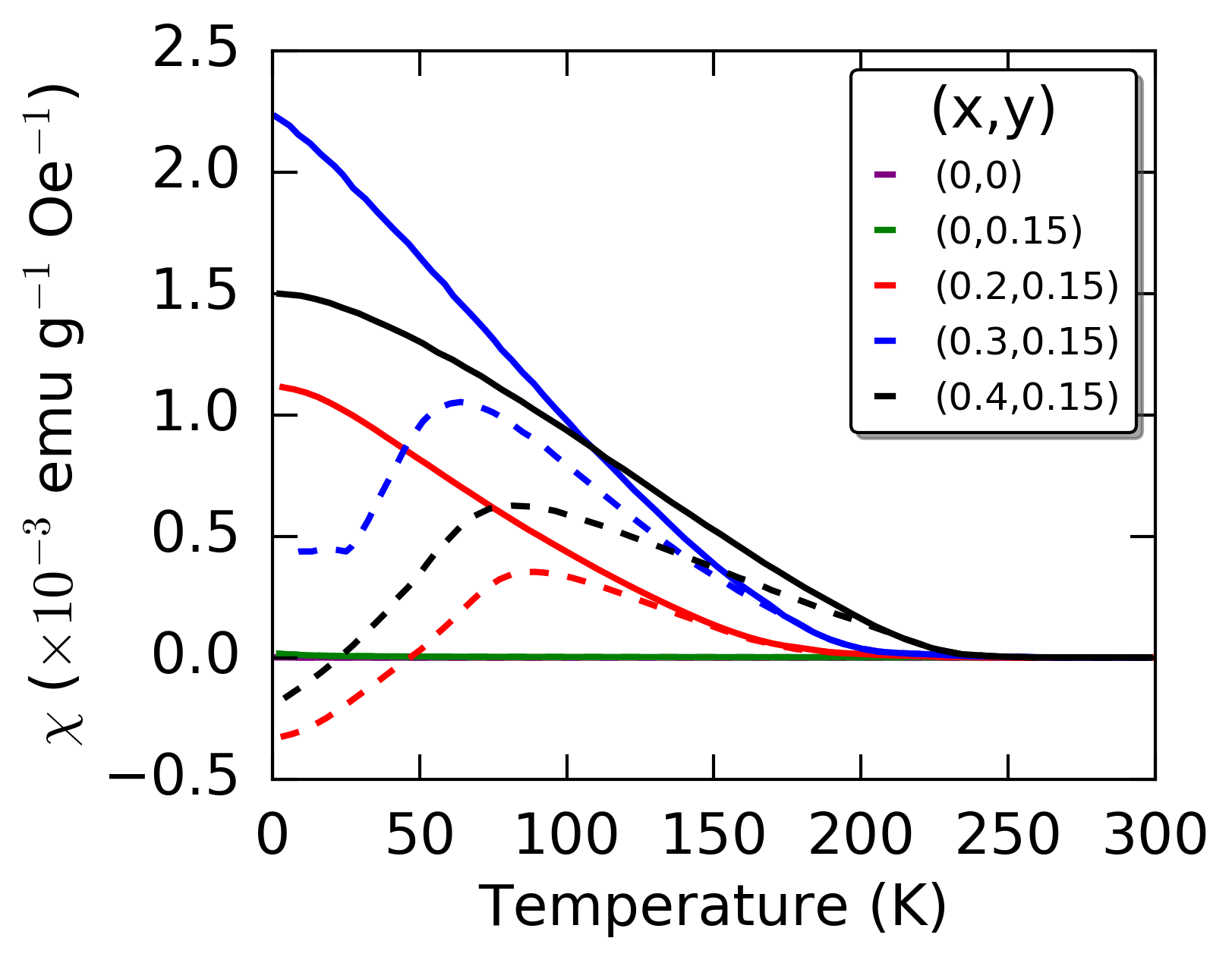}
		\caption{\label{fig:mag} (Color online) DC magnetic susceptibility of \bakxy. Solid (dashed) curves represent field-cooled (zero-field-cooled) measurements. A clear ferromagnetic response is seen for the samples with ($x,y$)= (0.2,~0.15), (0.3,~0.15), and (0.4,~0.15) around 180~K, 220~K, and 200~K, respectively.}
	\end{figure}

	The DC magnetic susceptibility $\chi$ for all five compositions of \bakxy\ is displayed in Fig.~\ref{fig:mag}. Measurements performed in a field-cooling protocol are shown as solid curves, and the dashed curves represent zero-field-cooled measurements. A clear upturn in the susceptibility is observed around 180~K, 220~K, and 200~K for the samples with ($x,y$)= (0.2,~0.15), (0.3,~0.15), and (0.4,~0.15), respectively, indicating the development of ferromagnetic order in these compounds. In contrast, the samples with no potassium content show no such ferromagnetic upturn of the susceptibility. These results are consistent with previously published data~\cite{zhao;nc13}.
	
\begin{figure}
		\includegraphics[width=80mm]{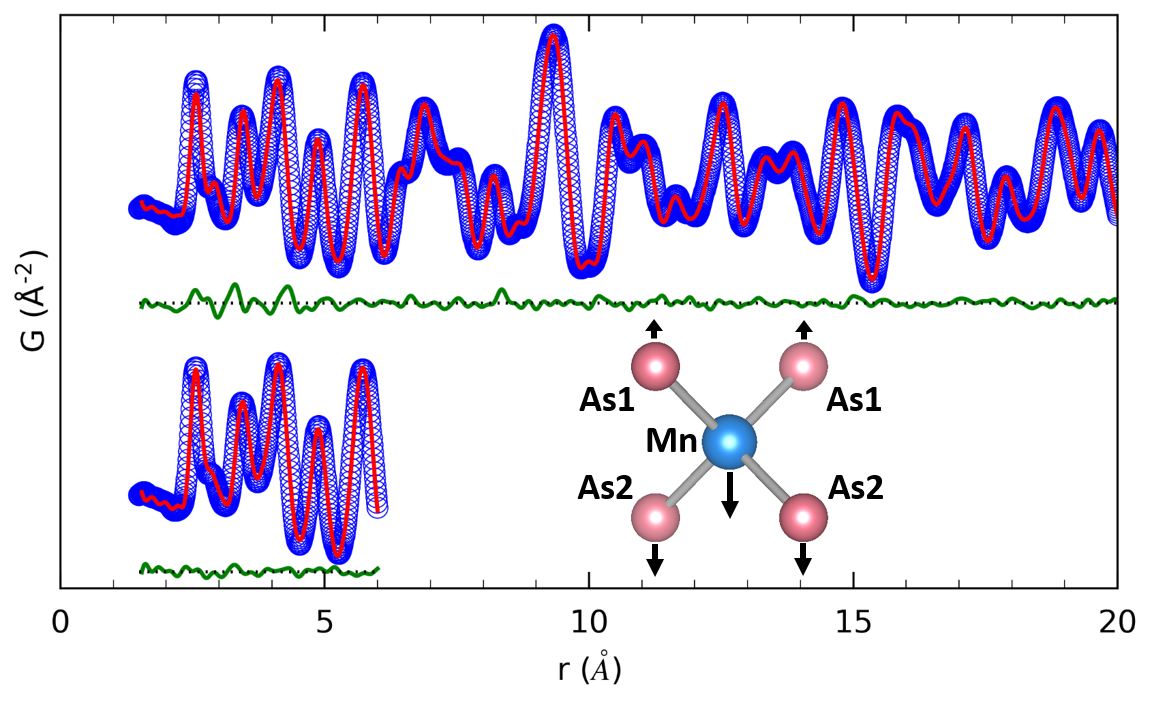}
		\caption{\label{fig:xray} (Color online) Atomic PDF fits for the $(x,y)=(0.3,0.15)$ sample at 300~K obtained from x-ray total scattering data. The blue curves represent experimental data, the red curves the calculated fit, and the green curves the residual, offset for clarity. The top set of curves comes from a fit from 1.5~\AA\ to 20~\AA\ using the tetragonal structure, and the lower set comes from a fit from 1.5~\AA\ to 6.0~\AA\ using the lower-symmetry orthorhombic model described in the main text. Inset: Close-up view of the MnAs$_4$ tetrahedral unit. The arrows indicate the displacements of the atoms in the distorted structure. The numerical labels are to be understood in reference to a MnAs$_4$ unit in the lower half of the unit cell displayed in Fig.~\ref{fig:struc}.}
	\end{figure}

	To investigate the behavior of the average and local atomic structure as the temperature is lowered through the magnetic transition temperature, we performed x-ray total scattering measurements of the ($x,y$)~=~(0.3,~0.15) sample at approximately 30 temperature points between 300~K and 100~K. In Fig.~\ref{fig:xray}, we display a representative fit to the atomic PDF at 300~K using the reported tetragonal crystal structure as the top set of curves. The fit residual, shown as the green curve offset below the experimental (blue) and calculated (red) curves, is quite small for $r \gtrsim 5$ \AA\ but much larger for smaller $r$, indicating that the expected tetragonal structure describes the data very well except for short-range atom pairs separated by less than about 5~\AA. Because it is so short-ranged, this local deviation from the average structure results in subtle diffuse scattering in the diffraction pattern and that would not be noticed in standard diffraction analysis, such as that published previously~\cite{zhao;nc13}. The fact that it is very easily observed in the PDF pattern highlights the power of real-space analysis of total scattering data.

	To gain a more detailed understanding of the local structure, we defined a lower-symmetry structural model and refined it against the PDF data from 1.5~\AA\ to 6.0~\AA. We made three modifications to the original tetragonal I4/mmm structure to arrive at our candidate lower-symmetry model: (1) We allowed displacement of the Zn/Mn site along the $c$-axis; (2) we allowed the $c$-axis displacements of the As sites above and below the Zn/Mn plane to be independent from each other; and (3) we allowed an orthorhombic lattice with $a \ne b$. In the resulting fit to the low-$r$ data (shown as the lower set of curves in Fig.~\ref{fig:xray} for the (0.3, 0.15) sample at 300~K), the residual curve is greatly diminished, indicating a much higher quality of fit. It is possible that a lower symmetry structure would fit the data even better, but at the expense of introducing additional free parameters. The fit shown was achieved by introducing only three additional parameters, and it appears to be the simplest model capable of significantly improving upon the tetragonal model. We note that this lower-symmetry structure is highly local: the orthorhombic fits rapidly deteriorate in quality for larger $r$.
	
\begin{figure}
		\includegraphics[width=80mm]{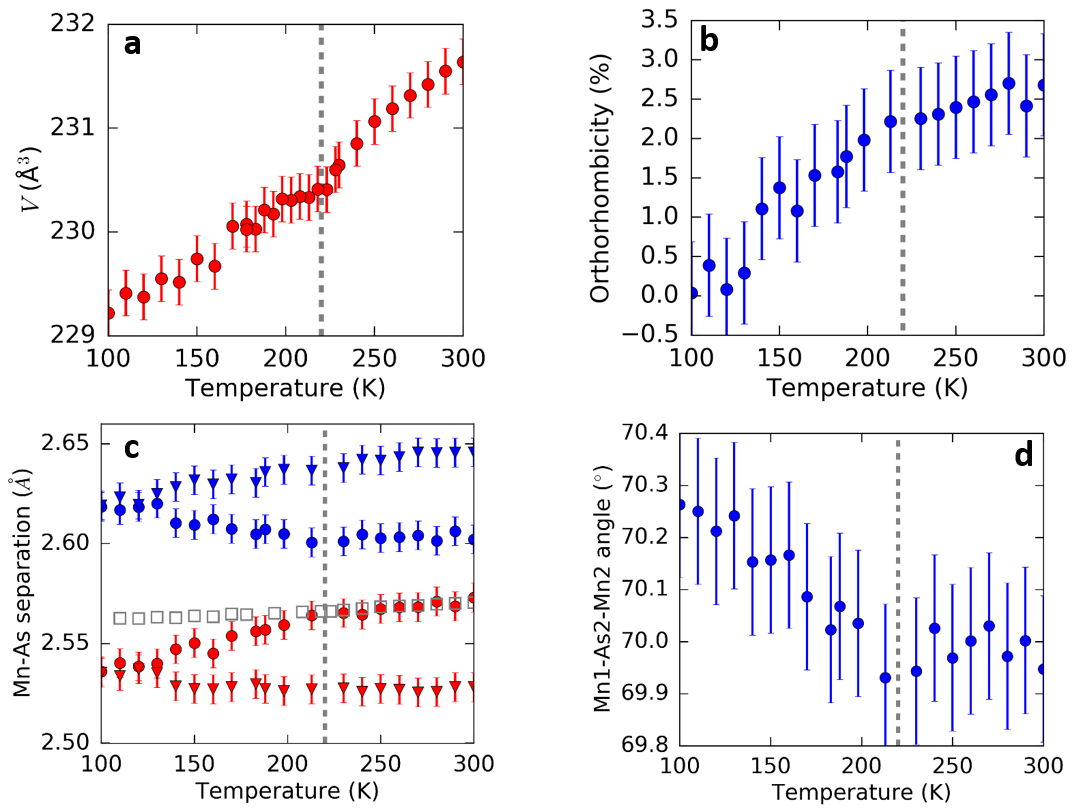}
		\caption{\label{fig:positions} (Color online) Results of structural refinements of Ba$_{0.7}$K$_{0.3}$(Zn$_{0.85}$Mn$_{0.15}$)$_2$As$_2$. (a) Temperature dependence of the unit cell volume from the tetragonal model. (b) Orthorhombicity (defined in main text) determined from refinements using the orthorhombic model. (c) Separation distance of the four distinct Mn-As bonds in the local orthorhombic structure. Based on the labels in Fig.~\ref{fig:struc}a, these bonds are Mn1-As1 (blue triangles), Mn2-As1 (blue circles), Mn1-As2 (red circles) and Mn2-As2 (red triangles). The open gray squares show the Mn-As distance in the tetragonal model. (d) Angle formed by the Mn1-As2-Mn2 chain of atoms in the orthorhombic model. The error bars represent the estimated standard deviation of the parameters obtained from the least-squares refinements.}
	\end{figure}

	Using the tetragonal model for fits up to 30~\AA\ and the orthorhombic up to 6~\AA, we performed fits to the $\sim$30 PDFs for the ($x,y$)=(0.3,~0.15) sample and extracted the refined parameters. The temperature dependences of several of the refined parameters from these models are shown in Fig.~\ref{fig:positions}. Panel (a) displays the unit cell volume for the tetragonal model. Below $\sim$250~K, the volume begins to drop below its linear high-temperature trend. At 220~K (the magnetic ordering temperature for this compound), a kink occurs in the volume, with the volume decreasing at a slower rate as the temperature is lowered further. Below $\sim$200~K, the volume recovers an approximately linear temperature dependence similar to the high-temperature trend. The slightly jagged behavior of the unit cell volume at low temperature is a fitting artifact due to a very small, oscillating drift in the x-ray wavelength at the beamline.
	
	 Refinements using the orthorhombic model revealed a rather significant orthorhombic distortion at room temperature ($\sim \pm 0.06$~\AA\ from the average lattice parameter length). The mean of $a$ and $b$ is nearly identical to the tetragonal $a$ lattice parameter, with a difference of about 0.05\%. In Fig.~\ref{fig:positions}b, we display the temperature dependence of the orthorhombicity parameter $\eta$, defined as the ratio of the difference between $a$ and $b$ to the mean of $a$ and $b$, i.e. $\eta = 2(a-b)/(a+b)$. The orthorhombicity is largest at 300~K and decreases as the temperature is lowered, until the lattice parameters are restored to their tetragonal degeneracy ($\eta=0$) at 100~K. Similar to the lattice parameters and unit cell volume in the tetragonal model, the orthorhombicity shows a kink around the magnetic ordering temperature of 220~K; as the temperature is lowered below \Tc, the orthorhombicity decreases at a faster rate. The unit cell volume for the orthorhombic model (not shown) displays the same upward kink around 220~K as the tetragonal unit cell. The orthorhombic unit cell is systematically larger than the tetragonal unit cell by approximately 0.25\%.
	
	We also extracted the refined atomic positions from the lower-symmetry model, which are likely to have relevance for the microscopic physics at play in the system. For clarity, we have labeled particular atoms in Fig.~\ref{fig:struc}a as Mn1, Mn2, As1, and As2. The sites labeled Mn1 and Mn2 are of course highly likely to host a Zn atom instead of Mn, but for simplicity we refer to the sites merely as Mn1 and Mn2. At all temperatures measured, the $z$-coordinate of the Mn1 and Mn2 atoms is approximately 0.1~\AA\ lower than in the case of the tetragonal model, where it is constrained to have a fractional z-coordinate of 0.25. The As2 $z$-coordinate is also shifted downward by about 0.05~\AA\ relative to the tetragonal structure, while the As1 $z$-coordinate is shifted very slightly upward ($\sim$0.01~\AA) until about 130~K, where it intersects the corresponding $z$-coordinate of the tetragonal structure. These displacements are illustrated schematically in the inset of Fig.~\ref{fig:xray}. From these refined atomic positions, we determined the nearest-neighbor Mn-As bond lengths and Mn-As-Mn bond angle, both displayed in Fig.~\ref{fig:positions}c-d.
	Due to the orthorhombic distortion and independent displacements of the As1 and As2 atoms, the fourfold degeneracy of the Mn-As bond in the tetragonal model is fully lifted. Panel (c) of Fig.~\ref{fig:positions} displays all four of these bond lengths: Mn1-As1 (blue triangles), Mn2-As1 (blue circles), Mn1-As2 (red circles) and Mn2-As2 (red triangles). The Mn-As bond length for the tetragonal model is shown as open gray squares for reference. At 300~K, where the structure is most orthorhombic, the four bond lengths are well separated, pointing to a rather asymmetric local environment for the Mn and As atoms. As the orthorhombicity decreases, the four distinct bond lengths converge to two bond lengths, one separating the As1 atom from each Mn atom and the other between As2 and the Mn atoms. Because of the independent shifts of the As1 and As2 atoms, these two bond lengths remain significantly different ($\sim 0.08$~\AA) even when the $a$ and $b$ lattice parameters have converged at low temperature. The bond angle formed by the Mn1-As2-Mn2 chain is plotted in Fig.~\ref{fig:positions}d, where a slight upturn is seen to occur around 220~K as the temperature is lowered.  The Mn1-As1-Mn2 bond angle (not shown) does not display any pronounced temperature-dependent behavior and remains roughly constant at $\sim$67.7$^{\circ}$.
	
	In contrast to the ($x,y$)~=~(0.3,~0.15) sample with a well-defined local structure, the (0,~0) sample showed no evidence for short-range correlations differing from the average tetragonal structure. Even when we used the lower-symmetry orthorhombic model to fit from 1.5~\AA\ to 6.0~\AA, the $a$ and $b$ lattice parameters converged robustly to nearly identical values at all 14 temperatures measured between 100~K and 300~K.
	
	To check the reproducibility of the orthorhombic model for other dopings, we measured the (0,~0) and (0.2,~0.15) samples at 100~K at the NSLS-II. Once again, the undoped sample showed no evidence for a local distortion, while the doped sample showed an improved fit below 6~\AA\ with the orthorhombic structure. The orthorhombic distortion and atomic displacements were comparable to the original results for (0.3,~0.15) obtained at the NSLS.
	
\begin{figure}
		\includegraphics[width=80mm]{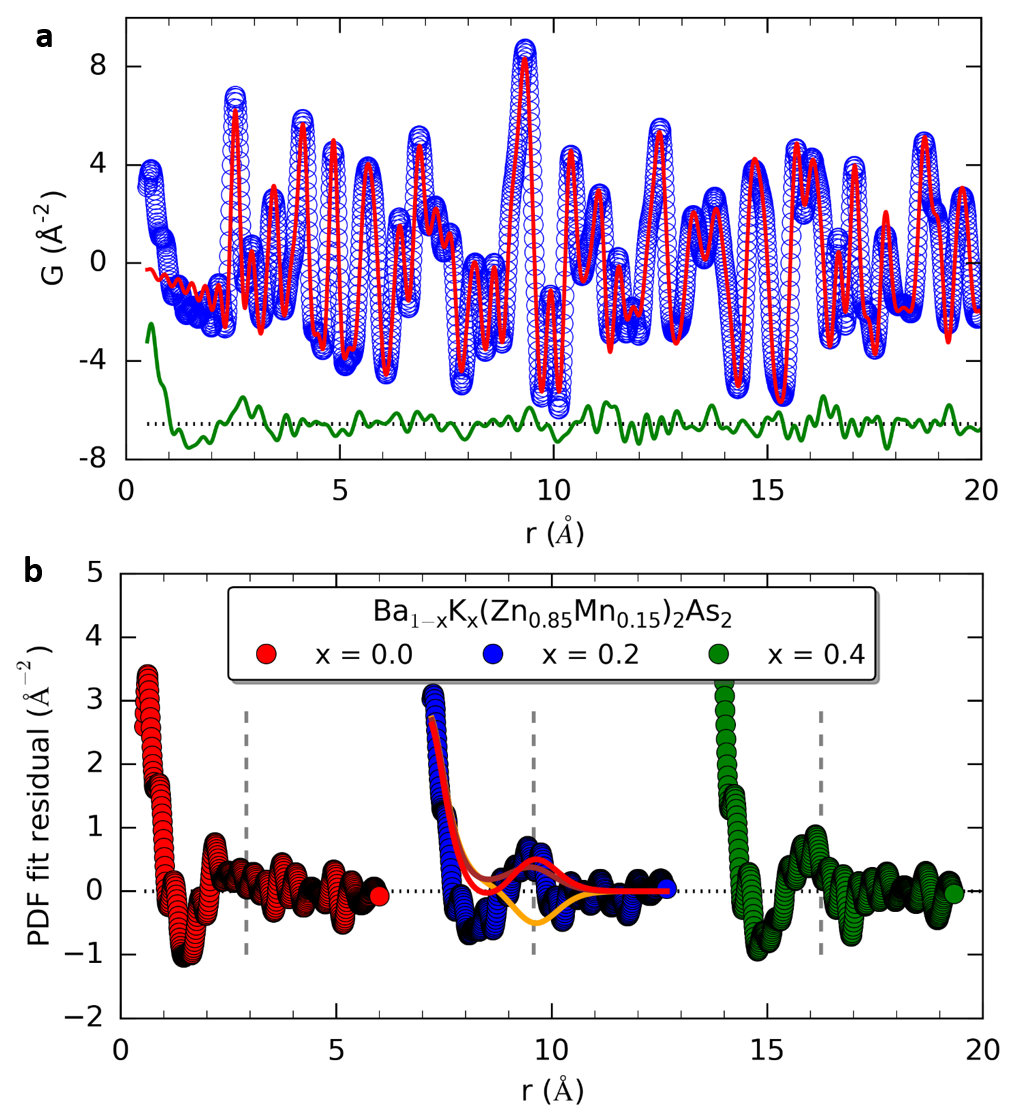}
		\caption{\label{fig:neutron} (Color online) Neutron PDF fits to \bakxy. (a) Refinement of the tetragonal model (red curve) against the 2~K data for the sample with ($x,y$) = (0.2,~0.15) (blue circles), with the fit residual shown below in green. (b) Nuclear PDF fit residuals for the (0,~0.15) (left, red), (0.2,~0.15) (middle, blue), and (0.4,~0.15) (right, green) samples. The latter two are shifted horizontally for clarity. The vertical dashed lines represent the Mn-Mn NN distance. The solid colored curves for the $x=0.2$ sample represent mPDF calculations described in the main text.}
	\end{figure}

	We also performed neutron time-of-flight total scattering measurements on the samples with ($x,y$)~=~(0,~0.15), (0.2,~0.15), and (0.4,~0.15) to further verify the local orthorhombic model and attempt to gain insight into the magnetic structure. The (0.2,~0.15) sample was measured at 7 temperatures between 300~K and 2~K, and the others were measured at 2~K and just one or two higher temperatures. Fig.~\ref{fig:neutron}a displays a fit to the (0.2,~0.15) sample data at 2~K using the tetragonal structure. The overall noise level in the data is significantly higher than the x-ray data due to background contributions from the neutron beamline, making it more difficult to visually identify any evidence for a local orthorhombic distortion. Nevertheless, fits to the low-$r$ region of the data (0.55~\AA\ to 6.0~\AA) using the orthorhombic model result in the same temperature-dependent trend in the orthorhombic lattice parameters as found in the x-ray data: $a$ and $b$ are well separated at 300~K, approach each other on cooling, and become nearly identical to each other below 100~K. 

	Because neutrons scatter off both nuclei and magnetic moments in a material, the total PDF also contains the Fourier transform of the magnetic scattering, i.e. the mPDF\cite{frand;aca14,frand;aca15,frand;prl16}. When the total scattering data are processed according to standard protocols for the atomic PDF, the magnetic scattering is not normalized by the magnetic form factor (as would otherwise be done to obtain a proper mPDF), so the mPDF contribution is broadened out by a factor roughly equal to $\sqrt{2}$ times the width of the real-space extent of a single localized magnetic moment~\cite{frand;aca15}. Hence, the mPDF appears as a broad signal underneath the much sharper atomic PDF peaks. Additionally, the unnormalized mPDF contains a peak at very low $r$ that arises as a consequence of not having normalized the magnetic scattering by the magnetic form factor before performing the Fourier transform, rather than from pairwise magnetic correlations.
	
	Since a model of the nuclear PDF will not capture these features of the mPDF, the experimental mPDF is often viewed most easily by examining the fit residual after performing a refinement of the nuclear structure against the total PDF signal~\cite{frand;aca15}. Indeed, upon inspection of the residual in Fig.~\ref{fig:neutron}a, we observe features that are consistent with these signatures of the mPDF: A large downward slope for $r \lesssim 1.3$~\AA\ arising from the tail end of the low-$r$ mPDF peak, and a broad positive peak centered around the Mn-Mn nearest neighbor (NN) distance of 2.91~\AA, consistent with ferromagnetic alignment of NN spins. If the NN correlations were antiferromagnetic, then the peak would be negative, but it is positive consistent with the expected ferromagnetic correlations.
	
Features in the PDF at very low values of $r$ can have non-physical origins coming from imperfectly corrected experimental aberrations in the data~\cite{egami;b;utbp12}.  However, already  by the nearest-neigbor interatomic distance these aberrations are small and hardly distort the PDF.  With
this in mind, the very low-$r$ peak $r \lesssim 1.3$~\AA\ is likely to contain non-magnetic contributions.  However, the positive correlation peak centered at 2.91~\AA\ is expected to be of magnetic origin.	To increase our confidence that this is the case, we compare the nuclear structure fit residuals (nominally the mPDF signal) for all three concentrations measured with neutrons. This is seen in Fig.~\ref{fig:neutron}b where the red symbols represent the first 6~\AA\ of the ($x,y$) = (0,~0.15) fit, the blue curve represents the (0.2,~0.15) fit residual, and the green curve the (0.4,~0.15) residual. The latter two are shifted in $r$ for easy side-by-side comparison. As a reminder, the magnetic susceptibility measurements showed the (0,~0.15) sample to be non-ferromagnetic and the other two to be ferromagnetic. The PDF fit residuals for all three patterns have a similar initial downward slope at very low $r$, which is expected given that all three samples have the same Mn content and therefore moment, though as discussed above, may also contain aberrations. Moving to the nearest neighbor correlation peak, the (0,~0.15) sample shows no clear feature at the Mn-Mn NN distance (indicated by the vertical dashed line), whereas the (0.2,~0.15) sample shows a well-defined peak at that distance, and the (0.4,~0.15) sample an even larger one. This suggests that no appreciable Mn-Mn spin correlations exist in the (0,~0.15) sample but they do in the other two compositions, consistent with our expectations from the magnetization data and our general understanding of hole-mediated ferromagnetism in DFS systems---with no K content, the (0,~0.15) sample has no itinerant holes to ferromagnetically couple the Mn spins.
	
	By comparing the calculated mPDF from various NN spin configurations to the observed peak in the data, we can gain insight into the details of the spin orientation in the material. The solid lines overlaid on the data for the (0.2,~0.15) sample in Fig.~\ref{fig:neutron}b represent the calculated mPDF from three distinct configurations of an arbitrary Mn spin and its four nearest neighbors (all assumed to be Mn for these simulations): All spins aligned along the $c$ axis (red curve), all spins aligned along the $a$ axis (brown curve), and the central spin anti-aligned with its four neighbors along the $c$-axis (orange curve). The configuration with all spins aligned along the $c$ axis clearly matches the data best. This is also consistent with recent single-crystal magnetization measurements which show a strong out-of-plane magnetization component and a much weaker in-plane component.
	
\begin{figure}
			\includegraphics[width=80mm]{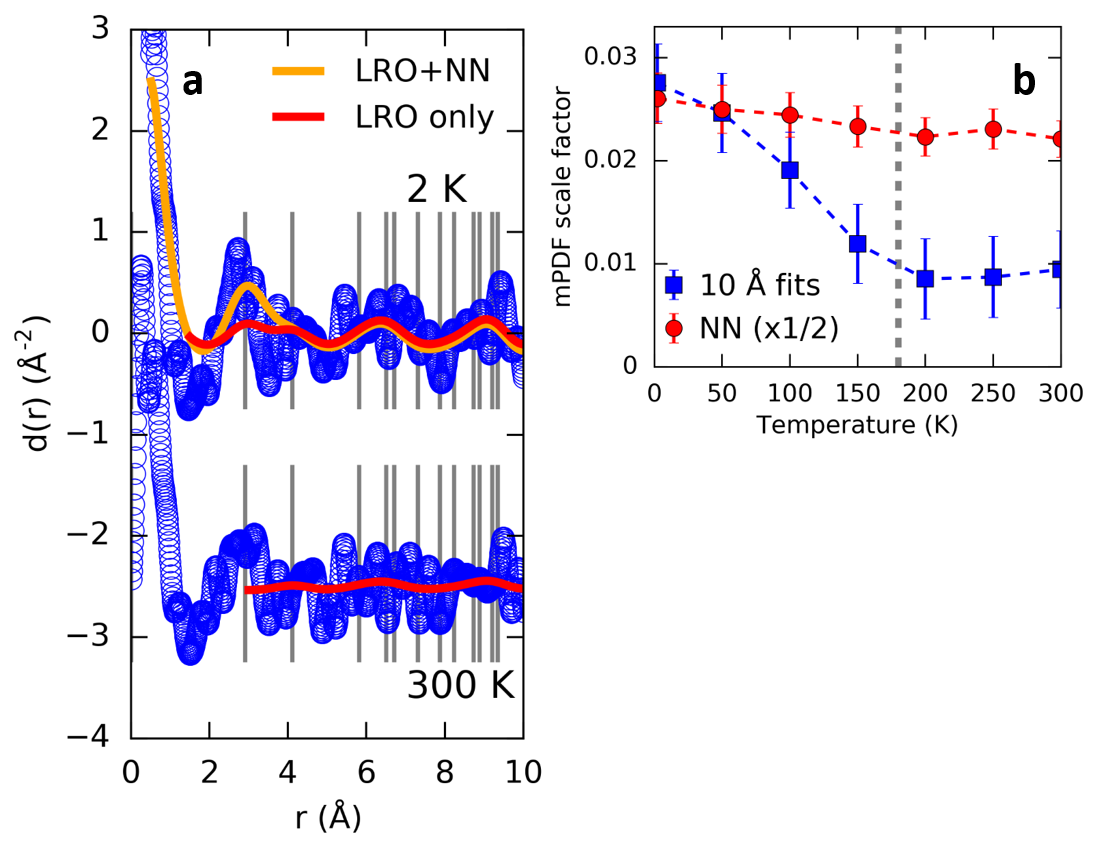}
			\caption{\label{fig:mPDFscale} (Color online) (a) Experimental mPDF patterns (blue curves) obtained as fit residuals from the nuclear PDF for 2~K (top) and 300~K (bottom), with calculated mPDFs from various models described in the text shown as red and orange curves. Mn-Mn bond lengths are shown as gray vertical lines. (b) mPDF scale factors from refinements of two models described in the text, with \Tc=180~K indicated by the dashed vertical line. The error bars represent the estimated standard deviation of the parameters obtained from the least-squares refinements.}
		\end{figure}

	Longer-range magnetic correlations among Mn spins beyond the first nearest neighbor would also be expected to contribute to the mPDF signal for the ferromagnetic samples. Using a model in which spins on every Zn/Mn site exhibit long-range ferromagnetic alignment (thus implicitly assuming a uniform distribution of Mn ions), we refined the mPDF scale factor and spin direction against the first 10~\AA\ of the nuclear PDF fit residual. For unnormalized experimental mPDF patterns from ferromagnetic materials obtained through standard PDF data reduction protocols, an additional linear correction proportional to the volume density of spins and the ordered ferromagnetic moment must be included to ensure a good fit~\cite{frand;unpub16}. The refined mPDFs from this model for data collected from the (0.2,~0.15) sample at 2~K and 300~K are displayed as the red curves (labeled ``LRO only'' to indicate the long-range-ordered model of ferromagnetism) in Fig~\ref{fig:mPDFscale}(a), with the Mn-Mn distances displayed as vertical gray lines. For the 2~K data, the calculated mPDF clearly captures the broad positive peaks around 4~\AA, 6.5~\AA, and 9~\AA, reflecting the ferromagnetic alignment of the Mn spins. The spin direction refined robustly to lie along the $c$-axis, in agreement with the earlier analysis of the NN mPDF peak. The negative features around 5~\AA\ and 8~\AA, which one might think would suggest antiferromagnetic correlations, are actually not due to magnetic correlations at all; this can be recognized by the fact that there are no Mn-Mn bonds near 5~\AA\ and significantly fewer bonds around 8~\AA\ than at the distances with clear positive peaks. Instead, these negative features are a consequence of the downward-sloping linear correction term introduced during the data reduction process.
	
	Interestingly, the calculated mPDF severely undershoots the initial positive peak at around 2.91~\AA\ arising from the NN spins. This indicates that the ferromagnetic correlations are significantly stronger between NN spins than more distant spins. To model this, we included an additional mPDF contribution containing only NN spins. The result for 2~K is shown as the orange curve (labeled ``LRO+NN'') in Fig.~\ref{fig:mPDFscale}(a), which accurately captures the NN peak as well as the features at higher $r$. The refined mPDF scale factor for the NN component is about twice as large as that for the long-range model. Beyond 10~\AA, the noise in the data begins to exceed the mPDF signal, making quantitative analysis difficult. However, the fit was improved by including an overall Gaussian damping envelope with a best-fit width of 54~$\pm$~11~\AA, suggesting a rather short magnetic correlation length even at 2~K.
	
	The mPDF data collected at 300~K along with the best-fit mPDF from the long-range ferromagnetic model are displayed as the lower set of curves in Fig.~\ref{fig:mPDFscale}(a). The features at higher $r$ are greatly diminished and perhaps entirely absent at 300~K compared to 2~K, as expected due to the disappearance of long-range magnetic order above \Tc=180~K. However, the mPDF scale factor converges to a nonzero value, suggesting that some correlations may survive out to 10~\AA, although this could also be due to random statistical noise or background contributions in the data. In either case, a clear peak persists at the NN distance of 2.91~\AA, indicating that rather significant ferromagnetic correlations among NN spins exist at least up to 300~K.
	
	In Fig.~\ref{fig:mPDFscale}(b), we display the temperature dependence of the refined mPDF scale factor for two models: the long-range ferromagnetic model for fits up to 10~\AA\ (blue squares), and a model containing only the NN spins for fits up to 4~\AA\ (red circles, multiplied by 1/2 for clarity). The longer-range model shows a significant upturn around 180~K on cooling in an order-parameter-like fashion, as expected for the development of long-range magnetic order. On the other hand, the scale factor for the NN model remains quite large at all temperatures and shows only a modest increase below 180~K, indicating that the NN ferromagnetic correlations remain relatively strong at all temperature measured.
		
	\section{Discussion}
	The three main findings introduced in this paper are (1) the observation of a response in the atomic structure around \Tc; (2) the existence of very a short-range, symmetry-lowering structural distortion present in both the ferromagnetic and paramagnetic phases; and (3) a robust ferromagnetic coupling between NN Mn spins persisting up to 300~K, with ferromagnetic correlations for more distant neighbors disappearing above \Tc. 
	
	With regard to the average structure, the kink in the tetragonal lattice parameters around the magnetic ordering temperature is consistent with the well-known phenomenon of volume magnetostriction observed in other ferromagnets such as nickel or iron. By extrapolating the linear trend of the volume above 250~K for the (0.3, 0.15) sample, we can estimate the magnetostriction quantity $\Delta V/V$, where $\Delta V$ is the difference between the observed and extrapolated volume, to be approximately $-1\times 10^{-3}$ at \Tc. This has the same sign and order of magnitude as ferromagnetic nickel,\cite{dutre;b;mtaaom93} indicating relatively strong magnetoelastic coupling in \bak.
	
	With regard to the local orthorhombic structure, its robust presence in the doped samples and absence in the parent compound BaZn$_2$As$_2$ compound suggests that the impurity atoms induce local lattice distortions. The distorted structure involves significant displacements of the Zn/Mn and As atoms, suggesting that the Mn impurities play a more important role in the local distortion than the K impurities. Indeed, Mn impurities are known to cause local distortions of the MnAs$_4$ tetrahedra in other DFS systems such as (Ga,Mn)As and (Ga,Mn)N.~\cite{sapeg;prb02,luo;prb05} Our results indicate that the MnAs$_4$ tetrahedra in \bak\ are likewise distorted locally, with the MnAs$_4$ tetrahedra undergoing an overall expansion.
	
	Steric effects associated with the different sizes of the Zn and Mn ions may contribute to this local distortion. To investigate this possibility, we calculated the bond valence sums (BVS's) for the Zn$^{2+}$ and Mn$^{2+}$ ions using the refined atomic positions from the average tetragonal model and the local orthorhombic model. For the (0.3, 0.15) sample at 300~K, the Zn BVS's are 1.64 and 1.59 for the tetragonal and orthorhombic models, respectively; for Mn, the BVS's are 2.27 and 2.20. The significant difference between the Zn and Mn BVS's indeed suggests that different local environments would be expected around Zn and Mn ions. A plausible interpretation is that the structure relaxes locally around the Mn and Zn atoms such that the Mn-As and Zn-As bonds can approach their ideal lengths~\cite{jeong;prb01}. For Mn, which has a slightly larger atomic radius than Zn, this results in a net expansion of the MnAs$_4$ tetrahedra relative to the ZnAs$_4$ tetrahedra, albeit in an asymmetrical fashion. After periodic averaging (i.e. in the crystallographic picture), the refined lattice parameters and bond lengths converge to some average value intermediate between the shorter Zn-As bonds and the longer Mn-As bonds, with slightly enlarged atomic displacement parameters. The different local bonding environments of Zn and Mn likely contribute to the observed local symmetry lowering.
	
	To check for the possibility of phase segregation into Zn-rich and Mn-rich regions, we performed fits with a two-phase model, one phase solely with Zn and one solely with Mn. This two-phase model yielded an equally good fit as the single-phase orthorhombic model when restricted to 6~\AA\ and below, but for fits performed over longer $r$-ranges, the single-phase tetragonal model fit just as well or better, suggesting that there is no tendency for Zn and Mn segregation on length scales longer than approximately 5~\AA. This is consistent with previous neutron diffraction and muon spin relaxation data, which also found no evidence for an inhomogeneous distribution of Mn ions.
	
	We now discuss the mPDF results, which revealed a strong ferromagnetic peak at the Mn-Mn NN distance and somewhat weaker ferromagnetic peaks at higher $r$ for the doped compounds, while the compound with no K content displayed no mPDF signal. The relatively weak longer-range mPDF signal in the doped samples is not necessarily surprising, considering that even the high-resolution, low-noise neutron diffraction measurements reported earlier~\cite{zhao;nc13} detected no signature of the ferromagnetic order that was seen by \musr~\cite{zhao;nc13}. The present mPDF findings may help reconcile the apparent inconsistency of the previous neutron and \musr\ results; as a probe of local magnetism, \musr\ is sensitive to even very short-range magnetic correlations such as the strong NN correlations revealed by mPDF, while standard analysis of magnetic Bragg peaks is only applicable to long-range magnetic order and is insensitive to short-range correlations. The persistence of the NN mPDF signal above \Tc, which is in contrast to the disappearance of the \musr\ signal above \Tc, suggests a change in time scale of the magnetic correlations: above \Tc, the correlations fluctuate on shorter time scales that render them invisible to \musr\ but still observable in the diffuse neutron diffraction signal and mPDF.
	
	The presence of robust ferromagnetic correlations between NN spins suggested by the neutron data is in contrast to the two first-principles studies performed on this system,~\cite{glasb;prb14,yang;sss15} which both predict antiferromagnetic coupling of NN spins due to short-range superexchange between neighboring Mn spins mediated by the As anions. One possible explanation for this discrepancy is that the Mn ions exist in a highly asymmetrical local environment, with significantly different bond lengths and angles formed by the neighboring As anions. This could reduce the efficiency of the antiferromagnetic superexchange and allow the hole-mediated ferromagnetic exchange or some other mechanism to dominate instead. To our knowledge, the previous studies assumed the long-range tetragonal structure reported previously,~\cite{zhao;nc13} and so would not have included any of the possible effects of the lower local symmetry reported in the present paper. A calculation of the magnetic interactions given the local distortion revealed by the PDF data is beyond the scope of this paper but would be a worthwhile endeavor.
	
	The theoretical studies further suggest that the As anions could also become spin-polarized.~\cite{glasb;prb14,yang;sss15} However, the expectation is that the As magnetic moments would be very small compared to the Mn spins. Since the magnitude of the mPDF contribution of a pair of spins is proportional to the product of the two spins, contributions from As moments would be difficult to detect in the mPDF data.
	
	An experimental fact that may be difficult to reconcile with the picture of ferromagnetically coupled NN spins is that the saturated net magnetic moment tends to be $\sim$1-2~$\mu_{\mathrm{B}}$ per Mn, significantly smaller than would be expected from a full contribution of all the $S=5/2$ spins. The formation of antiferromagnetic NN pairs has often been invoked to explain this discrepancy, with which the claim of ferromagnetic NN spin pairs is clearly not consistent. On the other hand, there are other mechanisms that could also reduce the local moment size. For instance, it is known from recent x-ray absorption and photoemission spectroscopy that the Mn 3$d$ orbitals strongly hybridize with the As 4$p$ orbitals, which can also significantly reduce the local moment. In addition, local structural distortions in (Ga,Mn)As have been predicted to reduce the Mn moment,~\cite{rana;ijph12} which could also be the case for the local distortion in \bak. Some combination of these and other effects may explain the observed reduced ferromagnetic moment while remaining consistent with our observation of significant Mn NN ferromagnetic coupling. Nevertheless, this puzzling situation requires further investigation.

	\section{Conclusion}
	We have performed PDF analysis of x-ray and neutron total scattering measurements conducted on several compositions of \bak. The atomic PDF analysis revealed that the average crystallographic structure retains its tetragonal structure at all temperatures, but the unit cell volume shows an anomaly at the ferromagnetic ordering temperature, consistent with volume magnetostriction. In addition, a well-defined local orthorhombic distortion with asymmetric displacements of the As and Zn/Mn atoms exists on a very short length scale of $\lesssim 5$~\AA. On cooling, the orthorhombic distortion gradually relaxes back to the average tetragonal structure, although the local environment of the Zn/Mn and As sites remains distorted. The neutron PDF data are consistent with this local orthorhombic distortion and also indicate robust short-range, ferromagnetic alignment of NN Mn spins along the $c$ axis that persists well above \Tc, along with somewhat weaker ferromagnetic correlations among further Mn neighbors that increase below the expected magnetic ordering temperature. The existence of a lower-symmetry local structure and ferromagnetic (rather than antiferromagnetic) coupling of NN Mn spins are both unexpected results that may have significant implications for the mechanism of ferromagnetism in this system.

	\begin{acknowledgments}
		We thank Joan Siewenie for technical assistance with the measurements performed on the NOMAD instrument. BAF and YJU acknowledge support from the NSF via PIRE program OISE-0968226 and DMREF program DMR-1436095, and BAF by the NSF GRFP
		program DGE-11-44155. SJLB acknowledges support from the U.S. Department of Energy, Office of Science, Office of Basic Energy Sciences (DOE-BES) under contract No. DE-SC00112704. Use of the National Synchrotron Light Source (NSLS) and the NSLS-II, at Brookhaven National Laboratory, was supported by DOE-BES under contract No. DE-SC0012704. Use of the Spallation Neutron Source, Oak Ridge National Laboratory, was sponsored by the Scientific User Facilities Division, Office of Basic Energy Science, U.S. DOE.
	\end{acknowledgments}
	

\begin{thebibliography}{29}%
	\makeatletter
	\providecommand \@ifxundefined [1]{%
		\@ifx{#1\undefined}
	}%
	\providecommand \@ifnum [1]{%
		\ifnum #1\expandafter \@firstoftwo
		\else \expandafter \@secondoftwo
		\fi
	}%
	\providecommand \@ifx [1]{%
		\ifx #1\expandafter \@firstoftwo
		\else \expandafter \@secondoftwo
		\fi
	}%
	\providecommand \natexlab [1]{#1}%
	\providecommand \enquote  [1]{``#1''}%
	\providecommand \bibnamefont  [1]{#1}%
	\providecommand \bibfnamefont [1]{#1}%
	\providecommand \citenamefont [1]{#1}%
	\providecommand \href@noop [0]{\@secondoftwo}%
	\providecommand \href [0]{\begingroup \@sanitize@url \@href}%
	\providecommand \@href[1]{\@@startlink{#1}\@@href}%
	\providecommand \@@href[1]{\endgroup#1\@@endlink}%
	\providecommand \@sanitize@url [0]{\catcode `\\12\catcode `\$12\catcode
		`\&12\catcode `\#12\catcode `\^12\catcode `\_12\catcode `\%12\relax}%
	\providecommand \@@startlink[1]{}%
	\providecommand \@@endlink[0]{}%
	\providecommand \url  [0]{\begingroup\@sanitize@url \@url }%
	\providecommand \@url [1]{\endgroup\@href {#1}{\urlprefix }}%
	\providecommand \urlprefix  [0]{URL }%
	\providecommand \Eprint [0]{\href }%
	\providecommand \doibase [0]{http://dx.doi.org/}%
	\providecommand \selectlanguage [0]{\@gobble}%
	\providecommand \bibinfo  [0]{\@secondoftwo}%
	\providecommand \bibfield  [0]{\@secondoftwo}%
	\providecommand \translation [1]{[#1]}%
	\providecommand \BibitemOpen [0]{}%
	\providecommand \bibitemStop [0]{}%
	\providecommand \bibitemNoStop [0]{.\EOS\space}%
	\providecommand \EOS [0]{\spacefactor3000\relax}%
	\providecommand \BibitemShut  [1]{\csname bibitem#1\endcsname}%
	\let\auto@bib@innerbib\@empty
	\bibitem [{\citenamefont {Dietl}\ and\ \citenamefont
		{Ohno}(2014)}]{dietl;rmp14}%
	\BibitemOpen
	\bibfield  {author} {\bibinfo {author} {\bibfnamefont {T.}~\bibnamefont
			{Dietl}}\ and\ \bibinfo {author} {\bibfnamefont {H.}~\bibnamefont {Ohno}},\
	}\href {\doibase 10.1103/RevModPhys.86.187} {\bibfield  {journal} {\bibinfo
		{journal} {Rev. Mod. Phys.}\ }\textbf {\bibinfo {volume} {86}},\ \bibinfo
	{pages} {187} (\bibinfo {year} {2014})}\BibitemShut {NoStop}%
\bibitem [{\citenamefont {Ohno}(1998)}]{ohno;s98}%
\BibitemOpen
\bibfield  {author} {\bibinfo {author} {\bibfnamefont {H.}~\bibnamefont
		{Ohno}},\ }\href@noop {} {\bibfield  {journal} {\bibinfo  {journal}
		{Science}\ }\textbf {\bibinfo {volume} {281}},\ \bibinfo {pages} {951}
	(\bibinfo {year} {1998})}\BibitemShut {NoStop}%
\bibitem [{\citenamefont {Deng}\ \emph {et~al.}(2011)\citenamefont {Deng},
	\citenamefont {Jin}, \citenamefont {Liu}, \citenamefont {Wang}, \citenamefont
	{Zhu}, \citenamefont {Feng}, \citenamefont {Chen}, \citenamefont {Yu},
	\citenamefont {Arguello}, \citenamefont {Goko}, \citenamefont {Ning},
	\citenamefont {Zhang}, \citenamefont {Wang}, \citenamefont {Aczel},
	\citenamefont {Munsie}, \citenamefont {Williams}, \citenamefont {Luke},
	\citenamefont {Kakeshita}, \citenamefont {Uchida}, \citenamefont {Higemoto},
	\citenamefont {Ito}, \citenamefont {Gu}, \citenamefont {Maekawa},
	\citenamefont {Morris},\ and\ \citenamefont {Uemura}}]{deng;nc11}%
\BibitemOpen
\bibfield  {author} {\bibinfo {author} {\bibfnamefont {Z.}~\bibnamefont
		{Deng}}, \bibinfo {author} {\bibfnamefont {C.}~\bibnamefont {Jin}}, \bibinfo
	{author} {\bibfnamefont {Q.}~\bibnamefont {Liu}}, \bibinfo {author}
	{\bibfnamefont {X.}~\bibnamefont {Wang}}, \bibinfo {author} {\bibfnamefont
		{J.}~\bibnamefont {Zhu}}, \bibinfo {author} {\bibfnamefont {S.}~\bibnamefont
		{Feng}}, \bibinfo {author} {\bibfnamefont {L.}~\bibnamefont {Chen}}, \bibinfo
	{author} {\bibfnamefont {R.}~\bibnamefont {Yu}}, \bibinfo {author}
	{\bibfnamefont {C.}~\bibnamefont {Arguello}}, \bibinfo {author}
	{\bibfnamefont {T.}~\bibnamefont {Goko}}, \bibinfo {author} {\bibfnamefont
		{F.}~\bibnamefont {Ning}}, \bibinfo {author} {\bibfnamefont {J.}~\bibnamefont
		{Zhang}}, \bibinfo {author} {\bibfnamefont {Y.}~\bibnamefont {Wang}},
	\bibinfo {author} {\bibfnamefont {A.}~\bibnamefont {Aczel}}, \bibinfo
	{author} {\bibfnamefont {T.}~\bibnamefont {Munsie}}, \bibinfo {author}
	{\bibfnamefont {T.}~\bibnamefont {Williams}}, \bibinfo {author}
	{\bibfnamefont {G.}~\bibnamefont {Luke}}, \bibinfo {author} {\bibfnamefont
		{T.}~\bibnamefont {Kakeshita}}, \bibinfo {author} {\bibfnamefont
		{S.}~\bibnamefont {Uchida}}, \bibinfo {author} {\bibfnamefont
		{W.}~\bibnamefont {Higemoto}}, \bibinfo {author} {\bibfnamefont
		{T.}~\bibnamefont {Ito}}, \bibinfo {author} {\bibfnamefont {B.}~\bibnamefont
		{Gu}}, \bibinfo {author} {\bibfnamefont {S.}~\bibnamefont {Maekawa}},
	\bibinfo {author} {\bibfnamefont {G.}~\bibnamefont {Morris}}, \ and\ \bibinfo
	{author} {\bibfnamefont {Y.}~\bibnamefont {Uemura}},\ }\href@noop {}
{\bibfield  {journal} {\bibinfo  {journal} {Nat. Commun.}\ }\textbf {\bibinfo
		{volume} {2}},\ \bibinfo {pages} {422} (\bibinfo {year} {2011})}\BibitemShut
{NoStop}%
\bibitem [{\citenamefont {Zhao}\ \emph {et~al.}(2013)\citenamefont {Zhao},
	\citenamefont {Deng}, \citenamefont {Wang}, \citenamefont {Han},
	\citenamefont {Zhu}, \citenamefont {Li}, \citenamefont {Liu}, \citenamefont
	{Yu}, \citenamefont {Goko}, \citenamefont {Frandsen}, \citenamefont {Liu},
	\citenamefont {Ning}, \citenamefont {Uemura}, \citenamefont {Dabkowska},
	\citenamefont {Luke}, \citenamefont {Luetkens}, \citenamefont {Morenzoni},
	\citenamefont {Dunsiger}, \citenamefont {Senyshyn}, \citenamefont {Boeni},\
	and\ \citenamefont {Jin}}]{zhao;nc13}%
\BibitemOpen
\bibfield  {author} {\bibinfo {author} {\bibfnamefont {K.}~\bibnamefont
		{Zhao}}, \bibinfo {author} {\bibfnamefont {Z.}~\bibnamefont {Deng}}, \bibinfo
	{author} {\bibfnamefont {X.~C.}\ \bibnamefont {Wang}}, \bibinfo {author}
	{\bibfnamefont {W.}~\bibnamefont {Han}}, \bibinfo {author} {\bibfnamefont
		{J.~L.}\ \bibnamefont {Zhu}}, \bibinfo {author} {\bibfnamefont
		{X.}~\bibnamefont {Li}}, \bibinfo {author} {\bibfnamefont {Q.~Q.}\
		\bibnamefont {Liu}}, \bibinfo {author} {\bibfnamefont {R.~C.}\ \bibnamefont
		{Yu}}, \bibinfo {author} {\bibfnamefont {T.}~\bibnamefont {Goko}}, \bibinfo
	{author} {\bibfnamefont {B.}~\bibnamefont {Frandsen}}, \bibinfo {author}
	{\bibfnamefont {L.}~\bibnamefont {Liu}}, \bibinfo {author} {\bibfnamefont
		{F.}~\bibnamefont {Ning}}, \bibinfo {author} {\bibfnamefont {Y.~J.}\
		\bibnamefont {Uemura}}, \bibinfo {author} {\bibfnamefont {H.}~\bibnamefont
		{Dabkowska}}, \bibinfo {author} {\bibfnamefont {G.~M.}\ \bibnamefont {Luke}},
	\bibinfo {author} {\bibfnamefont {H.}~\bibnamefont {Luetkens}}, \bibinfo
	{author} {\bibfnamefont {E.}~\bibnamefont {Morenzoni}}, \bibinfo {author}
	{\bibfnamefont {S.~R.}\ \bibnamefont {Dunsiger}}, \bibinfo {author}
	{\bibfnamefont {A.}~\bibnamefont {Senyshyn}}, \bibinfo {author}
	{\bibfnamefont {P.}~\bibnamefont {Boeni}}, \ and\ \bibinfo {author}
	{\bibfnamefont {C.~Q.}\ \bibnamefont {Jin}},\ }\href@noop {} {\bibfield
	{journal} {\bibinfo  {journal} {Nat. Commun.}\ }\textbf {\bibinfo {volume}
		{4}},\ \bibinfo {pages} {1442} (\bibinfo {year} {2013})}\BibitemShut
{NoStop}%
\bibitem [{\citenamefont {Chen}\ \emph {et~al.}(2014)\citenamefont {Chen},
	\citenamefont {Zhao}, \citenamefont {Deng}, \citenamefont {Han},
	\citenamefont {Zhu}, \citenamefont {Wang}, \citenamefont {Liu}, \citenamefont
	{Frandsen}, \citenamefont {Liu}, \citenamefont {Cheung}, \citenamefont
	{Ning}, \citenamefont {Munsie}, \citenamefont {Medina}, \citenamefont {Luke},
	\citenamefont {Carlo}, \citenamefont {Munevar}, \citenamefont {Uemura},\ and\
	\citenamefont {Jin}}]{chen;prb14}%
\BibitemOpen
\bibfield  {author} {\bibinfo {author} {\bibfnamefont {B.~J.}\ \bibnamefont
		{Chen}}, \bibinfo {author} {\bibfnamefont {K.}~\bibnamefont {Zhao}}, \bibinfo
	{author} {\bibfnamefont {Z.}~\bibnamefont {Deng}}, \bibinfo {author}
	{\bibfnamefont {W.}~\bibnamefont {Han}}, \bibinfo {author} {\bibfnamefont
		{J.~L.}\ \bibnamefont {Zhu}}, \bibinfo {author} {\bibfnamefont {X.~C.}\
		\bibnamefont {Wang}}, \bibinfo {author} {\bibfnamefont {Q.~Q.}\ \bibnamefont
		{Liu}}, \bibinfo {author} {\bibfnamefont {B.}~\bibnamefont {Frandsen}},
	\bibinfo {author} {\bibfnamefont {L.}~\bibnamefont {Liu}}, \bibinfo {author}
	{\bibfnamefont {S.}~\bibnamefont {Cheung}}, \bibinfo {author} {\bibfnamefont
		{F.~L.}\ \bibnamefont {Ning}}, \bibinfo {author} {\bibfnamefont {T.~J.~S.}\
		\bibnamefont {Munsie}}, \bibinfo {author} {\bibfnamefont {T.}~\bibnamefont
		{Medina}}, \bibinfo {author} {\bibfnamefont {G.~M.}\ \bibnamefont {Luke}},
	\bibinfo {author} {\bibfnamefont {J.~P.}\ \bibnamefont {Carlo}}, \bibinfo
	{author} {\bibfnamefont {J.}~\bibnamefont {Munevar}}, \bibinfo {author}
	{\bibfnamefont {Y.~J.}\ \bibnamefont {Uemura}}, \ and\ \bibinfo {author}
	{\bibfnamefont {C.~Q.}\ \bibnamefont {Jin}},\ }\href {\doibase
	10.1103/PhysRevB.90.155202} {\bibfield  {journal} {\bibinfo  {journal} {Phys.
			Rev. B}\ }\textbf {\bibinfo {volume} {90}},\ \bibinfo {pages} {155202}
	(\bibinfo {year} {2014})}\BibitemShut {NoStop}%
\bibitem [{\citenamefont {Deng}\ \emph {et~al.}(2013)\citenamefont {Deng},
	\citenamefont {Zhao}, \citenamefont {Gu}, \citenamefont {Han}, \citenamefont
	{Zhu}, \citenamefont {Wang}, \citenamefont {Li}, \citenamefont {Liu},
	\citenamefont {Yu}, \citenamefont {Goko}, \citenamefont {Frandsen},
	\citenamefont {Liu}, \citenamefont {Zhang}, \citenamefont {Wang},
	\citenamefont {Ning}, \citenamefont {Maekawa}, \citenamefont {Uemura},\ and\
	\citenamefont {Jin}}]{deng;prb13}%
\BibitemOpen
\bibfield  {author} {\bibinfo {author} {\bibfnamefont {Z.}~\bibnamefont
		{Deng}}, \bibinfo {author} {\bibfnamefont {K.}~\bibnamefont {Zhao}}, \bibinfo
	{author} {\bibfnamefont {B.}~\bibnamefont {Gu}}, \bibinfo {author}
	{\bibfnamefont {W.}~\bibnamefont {Han}}, \bibinfo {author} {\bibfnamefont
		{J.~L.}\ \bibnamefont {Zhu}}, \bibinfo {author} {\bibfnamefont {X.~C.}\
		\bibnamefont {Wang}}, \bibinfo {author} {\bibfnamefont {X.}~\bibnamefont
		{Li}}, \bibinfo {author} {\bibfnamefont {Q.~Q.}\ \bibnamefont {Liu}},
	\bibinfo {author} {\bibfnamefont {R.~C.}\ \bibnamefont {Yu}}, \bibinfo
	{author} {\bibfnamefont {T.}~\bibnamefont {Goko}}, \bibinfo {author}
	{\bibfnamefont {B.}~\bibnamefont {Frandsen}}, \bibinfo {author}
	{\bibfnamefont {L.}~\bibnamefont {Liu}}, \bibinfo {author} {\bibfnamefont
		{J.}~\bibnamefont {Zhang}}, \bibinfo {author} {\bibfnamefont
		{Y.}~\bibnamefont {Wang}}, \bibinfo {author} {\bibfnamefont {F.~L.}\
		\bibnamefont {Ning}}, \bibinfo {author} {\bibfnamefont {S.}~\bibnamefont
		{Maekawa}}, \bibinfo {author} {\bibfnamefont {Y.~J.}\ \bibnamefont {Uemura}},
	\ and\ \bibinfo {author} {\bibfnamefont {C.~Q.}\ \bibnamefont {Jin}},\ }\href
{\doibase 10.1103/PhysRevB.88.081203} {\bibfield  {journal} {\bibinfo
		{journal} {Phys. Rev. B}\ }\textbf {\bibinfo {volume} {88}},\ \bibinfo
	{pages} {081203} (\bibinfo {year} {2013})}\BibitemShut {NoStop}%
\bibitem [{\citenamefont {Ning}\ \emph {et~al.}(2014)\citenamefont {Ning},
	\citenamefont {Man}, \citenamefont {Gong}, \citenamefont {Zhi}, \citenamefont
	{Guo}, \citenamefont {Ding}, \citenamefont {Wang}, \citenamefont {Goko},
	\citenamefont {Liu}, \citenamefont {Frandsen}, \citenamefont {Uemura},
	\citenamefont {Luetkens}, \citenamefont {Morenzoni}, \citenamefont {Jin},
	\citenamefont {Munsie}, \citenamefont {Luke}, \citenamefont {Wang},\ and\
	\citenamefont {Chen}}]{ning;prb14}%
\BibitemOpen
\bibfield  {author} {\bibinfo {author} {\bibfnamefont {F.~L.}\ \bibnamefont
		{Ning}}, \bibinfo {author} {\bibfnamefont {H.}~\bibnamefont {Man}}, \bibinfo
	{author} {\bibfnamefont {X.}~\bibnamefont {Gong}}, \bibinfo {author}
	{\bibfnamefont {G.}~\bibnamefont {Zhi}}, \bibinfo {author} {\bibfnamefont
		{S.}~\bibnamefont {Guo}}, \bibinfo {author} {\bibfnamefont {C.}~\bibnamefont
		{Ding}}, \bibinfo {author} {\bibfnamefont {Q.}~\bibnamefont {Wang}}, \bibinfo
	{author} {\bibfnamefont {T.}~\bibnamefont {Goko}}, \bibinfo {author}
	{\bibfnamefont {L.}~\bibnamefont {Liu}}, \bibinfo {author} {\bibfnamefont
		{B.~A.}\ \bibnamefont {Frandsen}}, \bibinfo {author} {\bibfnamefont {Y.~J.}\
		\bibnamefont {Uemura}}, \bibinfo {author} {\bibfnamefont {H.}~\bibnamefont
		{Luetkens}}, \bibinfo {author} {\bibfnamefont {E.}~\bibnamefont {Morenzoni}},
	\bibinfo {author} {\bibfnamefont {C.~Q.}\ \bibnamefont {Jin}}, \bibinfo
	{author} {\bibfnamefont {T.}~\bibnamefont {Munsie}}, \bibinfo {author}
	{\bibfnamefont {G.~M.}\ \bibnamefont {Luke}}, \bibinfo {author}
	{\bibfnamefont {H.}~\bibnamefont {Wang}}, \ and\ \bibinfo {author}
	{\bibfnamefont {B.}~\bibnamefont {Chen}},\ }\href {\doibase
	10.1103/PhysRevB.90.085123} {\bibfield  {journal} {\bibinfo  {journal} {Phys.
			Rev. B}\ }\textbf {\bibinfo {volume} {90}},\ \bibinfo {pages} {085123}
	(\bibinfo {year} {2014})}\BibitemShut {NoStop}%
\bibitem [{\citenamefont {Zhao}\ \emph
	{et~al.}(2014{\natexlab{a}})\citenamefont {Zhao}, \citenamefont {Chen},
	\citenamefont {Deng}, \citenamefont {Han}, \citenamefont {Zhao},
	\citenamefont {Zhu}, \citenamefont {Liu}, \citenamefont {Wang}, \citenamefont
	{Frandsen}, \citenamefont {Liu}, \citenamefont {Cheung}, \citenamefont
	{Ning}, \citenamefont {Munsie}, \citenamefont {Medina}, \citenamefont {Luke},
	\citenamefont {Carlo}, \citenamefont {Munevar}, \citenamefont {Zhang},
	\citenamefont {Uemura},\ and\ \citenamefont {Jin}}]{zhao;jap14}%
\BibitemOpen
\bibfield  {author} {\bibinfo {author} {\bibfnamefont {K.}~\bibnamefont
		{Zhao}}, \bibinfo {author} {\bibfnamefont {B.~J.}\ \bibnamefont {Chen}},
	\bibinfo {author} {\bibfnamefont {Z.}~\bibnamefont {Deng}}, \bibinfo {author}
	{\bibfnamefont {W.}~\bibnamefont {Han}}, \bibinfo {author} {\bibfnamefont
		{G.~Q.}\ \bibnamefont {Zhao}}, \bibinfo {author} {\bibfnamefont {J.~L.}\
		\bibnamefont {Zhu}}, \bibinfo {author} {\bibfnamefont {Q.~Q.}\ \bibnamefont
		{Liu}}, \bibinfo {author} {\bibfnamefont {X.~C.}\ \bibnamefont {Wang}},
	\bibinfo {author} {\bibfnamefont {B.}~\bibnamefont {Frandsen}}, \bibinfo
	{author} {\bibfnamefont {L.}~\bibnamefont {Liu}}, \bibinfo {author}
	{\bibfnamefont {S.}~\bibnamefont {Cheung}}, \bibinfo {author} {\bibfnamefont
		{F.~L.}\ \bibnamefont {Ning}}, \bibinfo {author} {\bibfnamefont {T.~J.~S.}\
		\bibnamefont {Munsie}}, \bibinfo {author} {\bibfnamefont {T.}~\bibnamefont
		{Medina}}, \bibinfo {author} {\bibfnamefont {G.~M.}\ \bibnamefont {Luke}},
	\bibinfo {author} {\bibfnamefont {J.~P.}\ \bibnamefont {Carlo}}, \bibinfo
	{author} {\bibfnamefont {J.}~\bibnamefont {Munevar}}, \bibinfo {author}
	{\bibfnamefont {G.~M.}\ \bibnamefont {Zhang}}, \bibinfo {author}
	{\bibfnamefont {Y.~J.}\ \bibnamefont {Uemura}}, \ and\ \bibinfo {author}
	{\bibfnamefont {C.~Q.}\ \bibnamefont {Jin}},\ }\href {\doibase
	http://dx.doi.org/10.1063/1.4899190} {\bibfield  {journal} {\bibinfo
		{journal} {J. Appl. Phys.}\ }\textbf {\bibinfo {volume} {116}},\ \bibinfo
	{eid} {163906} (\bibinfo {year} {2014}{\natexlab{a}}),\
	http://dx.doi.org/10.1063/1.4899190}\BibitemShut {NoStop}%
\bibitem [{\citenamefont {Ding}\ \emph {et~al.}(2013)\citenamefont {Ding},
	\citenamefont {Man}, \citenamefont {Qin}, \citenamefont {Lu}, \citenamefont
	{Sun}, \citenamefont {Wang}, \citenamefont {Yu}, \citenamefont {Feng},
	\citenamefont {Goko}, \citenamefont {Arguello}, \citenamefont {Liu},
	\citenamefont {Frandsen}, \citenamefont {Uemura}, \citenamefont {Wang},
	\citenamefont {Luetkens}, \citenamefont {Morenzoni}, \citenamefont {Han},
	\citenamefont {Jin}, \citenamefont {Munsie}, \citenamefont {Williams},
	\citenamefont {D'Ortenzio}, \citenamefont {Medina}, \citenamefont {Luke},
	\citenamefont {Imai},\ and\ \citenamefont {Ning}}]{ding;prb13}%
\BibitemOpen
\bibfield  {author} {\bibinfo {author} {\bibfnamefont {C.}~\bibnamefont
		{Ding}}, \bibinfo {author} {\bibfnamefont {H.}~\bibnamefont {Man}}, \bibinfo
	{author} {\bibfnamefont {C.}~\bibnamefont {Qin}}, \bibinfo {author}
	{\bibfnamefont {J.}~\bibnamefont {Lu}}, \bibinfo {author} {\bibfnamefont
		{Y.}~\bibnamefont {Sun}}, \bibinfo {author} {\bibfnamefont {Q.}~\bibnamefont
		{Wang}}, \bibinfo {author} {\bibfnamefont {B.}~\bibnamefont {Yu}}, \bibinfo
	{author} {\bibfnamefont {C.}~\bibnamefont {Feng}}, \bibinfo {author}
	{\bibfnamefont {T.}~\bibnamefont {Goko}}, \bibinfo {author} {\bibfnamefont
		{C.~J.}\ \bibnamefont {Arguello}}, \bibinfo {author} {\bibfnamefont
		{L.}~\bibnamefont {Liu}}, \bibinfo {author} {\bibfnamefont {B.~A.}\
		\bibnamefont {Frandsen}}, \bibinfo {author} {\bibfnamefont {Y.~J.}\
		\bibnamefont {Uemura}}, \bibinfo {author} {\bibfnamefont {H.}~\bibnamefont
		{Wang}}, \bibinfo {author} {\bibfnamefont {H.}~\bibnamefont {Luetkens}},
	\bibinfo {author} {\bibfnamefont {E.}~\bibnamefont {Morenzoni}}, \bibinfo
	{author} {\bibfnamefont {W.}~\bibnamefont {Han}}, \bibinfo {author}
	{\bibfnamefont {C.~Q.}\ \bibnamefont {Jin}}, \bibinfo {author} {\bibfnamefont
		{T.}~\bibnamefont {Munsie}}, \bibinfo {author} {\bibfnamefont {T.~J.}\
		\bibnamefont {Williams}}, \bibinfo {author} {\bibfnamefont {R.~M.}\
		\bibnamefont {D'Ortenzio}}, \bibinfo {author} {\bibfnamefont
		{T.}~\bibnamefont {Medina}}, \bibinfo {author} {\bibfnamefont {G.~M.}\
		\bibnamefont {Luke}}, \bibinfo {author} {\bibfnamefont {T.}~\bibnamefont
		{Imai}}, \ and\ \bibinfo {author} {\bibfnamefont {F.~L.}\ \bibnamefont
		{Ning}},\ }\href {\doibase 10.1103/PhysRevB.88.041102} {\bibfield  {journal}
	{\bibinfo  {journal} {Phys. Rev. B}\ }\textbf {\bibinfo {volume} {88}},\
	\bibinfo {pages} {041102} (\bibinfo {year} {2013})}\BibitemShut {NoStop}%
\bibitem [{\citenamefont {Zhao}\ \emph
	{et~al.}(2014{\natexlab{b}})\citenamefont {Zhao}, \citenamefont {Chen},
	\citenamefont {Zhao}, \citenamefont {Yuan}, \citenamefont {Liu},
	\citenamefont {Deng}, \citenamefont {Zhu},\ and\ \citenamefont
	{Jin}}]{zhao;csb14}%
\BibitemOpen
\bibfield  {author} {\bibinfo {author} {\bibfnamefont {K.}~\bibnamefont
		{Zhao}}, \bibinfo {author} {\bibfnamefont {B.}~\bibnamefont {Chen}}, \bibinfo
	{author} {\bibfnamefont {G.}~\bibnamefont {Zhao}}, \bibinfo {author}
	{\bibfnamefont {Z.}~\bibnamefont {Yuan}}, \bibinfo {author} {\bibfnamefont
		{Q.}~\bibnamefont {Liu}}, \bibinfo {author} {\bibfnamefont {Z.}~\bibnamefont
		{Deng}}, \bibinfo {author} {\bibfnamefont {J.}~\bibnamefont {Zhu}}, \ and\
	\bibinfo {author} {\bibfnamefont {C.}~\bibnamefont {Jin}},\ }\href@noop {}
{\bibfield  {journal} {\bibinfo  {journal} {Chinese Sci. Bull.}\ }\textbf
	{\bibinfo {volume} {59}},\ \bibinfo {pages} {2524} (\bibinfo {year}
	{2014}{\natexlab{b}})}\BibitemShut {NoStop}%
\bibitem [{\citenamefont {Suzuki}\ \emph
	{et~al.}(2015{\natexlab{a}})\citenamefont {Suzuki}, \citenamefont {Zhao},
	\citenamefont {Shibata}, \citenamefont {Takahashi}, \citenamefont {Sakamoto},
	\citenamefont {Yoshimatsu}, \citenamefont {Chen}, \citenamefont
	{Kumigashira}, \citenamefont {Chang}, \citenamefont {Lin}, \citenamefont
	{Huang}, \citenamefont {Chen}, \citenamefont {Gu}, \citenamefont {Maekawa},
	\citenamefont {Uemura}, \citenamefont {Jin},\ and\ \citenamefont
	{Fujimori}}]{suzuk;prb15a}%
\BibitemOpen
\bibfield  {author} {\bibinfo {author} {\bibfnamefont {H.}~\bibnamefont
		{Suzuki}}, \bibinfo {author} {\bibfnamefont {K.}~\bibnamefont {Zhao}},
	\bibinfo {author} {\bibfnamefont {G.}~\bibnamefont {Shibata}}, \bibinfo
	{author} {\bibfnamefont {Y.}~\bibnamefont {Takahashi}}, \bibinfo {author}
	{\bibfnamefont {S.}~\bibnamefont {Sakamoto}}, \bibinfo {author}
	{\bibfnamefont {K.}~\bibnamefont {Yoshimatsu}}, \bibinfo {author}
	{\bibfnamefont {B.~J.}\ \bibnamefont {Chen}}, \bibinfo {author}
	{\bibfnamefont {H.}~\bibnamefont {Kumigashira}}, \bibinfo {author}
	{\bibfnamefont {F.-H.}\ \bibnamefont {Chang}}, \bibinfo {author}
	{\bibfnamefont {H.-J.}\ \bibnamefont {Lin}}, \bibinfo {author} {\bibfnamefont
		{D.~J.}\ \bibnamefont {Huang}}, \bibinfo {author} {\bibfnamefont {C.~T.}\
		\bibnamefont {Chen}}, \bibinfo {author} {\bibfnamefont {B.}~\bibnamefont
		{Gu}}, \bibinfo {author} {\bibfnamefont {S.}~\bibnamefont {Maekawa}},
	\bibinfo {author} {\bibfnamefont {Y.~J.}\ \bibnamefont {Uemura}}, \bibinfo
	{author} {\bibfnamefont {C.~Q.}\ \bibnamefont {Jin}}, \ and\ \bibinfo
	{author} {\bibfnamefont {A.}~\bibnamefont {Fujimori}},\ }\href {\doibase
	10.1103/PhysRevB.91.140401} {\bibfield  {journal} {\bibinfo  {journal} {Phys.
			Rev. B}\ }\textbf {\bibinfo {volume} {91}},\ \bibinfo {pages} {140401}
	(\bibinfo {year} {2015}{\natexlab{a}})}\BibitemShut {NoStop}%
\bibitem [{\citenamefont {Suzuki}\ \emph
	{et~al.}(2015{\natexlab{b}})\citenamefont {Suzuki}, \citenamefont {Zhao},
	\citenamefont {Zhao}, \citenamefont {Chen}, \citenamefont {Horio},
	\citenamefont {Koshiishi}, \citenamefont {Xu}, \citenamefont {Kobayashi},
	\citenamefont {Minohara}, \citenamefont {Sakai}, \citenamefont {Horiba},
	\citenamefont {Kumigashira}, \citenamefont {Gu}, \citenamefont {Maekawa},
	\citenamefont {Uemura}, \citenamefont {Jin},\ and\ \citenamefont
	{Fujimori}}]{suzuk;prb15b}%
\BibitemOpen
\bibfield  {author} {\bibinfo {author} {\bibfnamefont {H.}~\bibnamefont
		{Suzuki}}, \bibinfo {author} {\bibfnamefont {G.~Q.}\ \bibnamefont {Zhao}},
	\bibinfo {author} {\bibfnamefont {K.}~\bibnamefont {Zhao}}, \bibinfo {author}
	{\bibfnamefont {B.~J.}\ \bibnamefont {Chen}}, \bibinfo {author}
	{\bibfnamefont {M.}~\bibnamefont {Horio}}, \bibinfo {author} {\bibfnamefont
		{K.}~\bibnamefont {Koshiishi}}, \bibinfo {author} {\bibfnamefont
		{J.}~\bibnamefont {Xu}}, \bibinfo {author} {\bibfnamefont {M.}~\bibnamefont
		{Kobayashi}}, \bibinfo {author} {\bibfnamefont {M.}~\bibnamefont {Minohara}},
	\bibinfo {author} {\bibfnamefont {E.}~\bibnamefont {Sakai}}, \bibinfo
	{author} {\bibfnamefont {K.}~\bibnamefont {Horiba}}, \bibinfo {author}
	{\bibfnamefont {H.}~\bibnamefont {Kumigashira}}, \bibinfo {author}
	{\bibfnamefont {B.}~\bibnamefont {Gu}}, \bibinfo {author} {\bibfnamefont
		{S.}~\bibnamefont {Maekawa}}, \bibinfo {author} {\bibfnamefont {Y.~J.}\
		\bibnamefont {Uemura}}, \bibinfo {author} {\bibfnamefont {C.~Q.}\
		\bibnamefont {Jin}}, \ and\ \bibinfo {author} {\bibfnamefont
		{A.}~\bibnamefont {Fujimori}},\ }\href {\doibase 10.1103/PhysRevB.92.235120}
{\bibfield  {journal} {\bibinfo  {journal} {Phys. Rev. B}\ }\textbf {\bibinfo
		{volume} {92}},\ \bibinfo {pages} {235120} (\bibinfo {year}
	{2015}{\natexlab{b}})}\BibitemShut {NoStop}%
\bibitem [{\citenamefont {Glasbrenner}\ \emph {et~al.}(2014)\citenamefont
	{Glasbrenner}, \citenamefont {\ifmmode \check{Z}\else
		\v{Z}\fi{}uti\ifmmode~\acute{c}\else \'{c}\fi{}},\ and\ \citenamefont
	{Mazin}}]{glasb;prb14}%
\BibitemOpen
\bibfield  {author} {\bibinfo {author} {\bibfnamefont {J.~K.}\ \bibnamefont
		{Glasbrenner}}, \bibinfo {author} {\bibfnamefont {I.}~\bibnamefont {\ifmmode
			\check{Z}\else \v{Z}\fi{}uti\ifmmode~\acute{c}\else \'{c}\fi{}}}, \ and\
	\bibinfo {author} {\bibfnamefont {I.~I.}\ \bibnamefont {Mazin}},\ }\href
{\doibase 10.1103/PhysRevB.90.140403} {\bibfield  {journal} {\bibinfo
		{journal} {Phys. Rev. B}\ }\textbf {\bibinfo {volume} {90}},\ \bibinfo
	{pages} {140403} (\bibinfo {year} {2014})}\BibitemShut {NoStop}%
\bibitem [{\citenamefont {Yang}\ \emph
	{et~al.}(2015{\natexlab{a}})\citenamefont {Yang}, \citenamefont {Luo},\ and\
	\citenamefont {Xiong}}]{yang;sss15}%
\BibitemOpen
\bibfield  {author} {\bibinfo {author} {\bibfnamefont {J.}~\bibnamefont
		{Yang}}, \bibinfo {author} {\bibfnamefont {S.}~\bibnamefont {Luo}}, \ and\
	\bibinfo {author} {\bibfnamefont {Y.}~\bibnamefont {Xiong}},\ }\href@noop {}
{\bibfield  {journal} {\bibinfo  {journal} {Solid State Sci.}\ }\textbf
	{\bibinfo {volume} {46}},\ \bibinfo {pages} {102 } (\bibinfo {year}
	{2015}{\natexlab{a}})}\BibitemShut {NoStop}%
\bibitem [{\citenamefont {Neuefeind}\ \emph {et~al.}(2012)\citenamefont
	{Neuefeind}, \citenamefont {Feygenson}, \citenamefont {Carruth},
	\citenamefont {Hoffmann},\ and\ \citenamefont {Chipley}}]{neufe;nimb12}%
\BibitemOpen
\bibfield  {author} {\bibinfo {author} {\bibfnamefont {J.}~\bibnamefont
		{Neuefeind}}, \bibinfo {author} {\bibfnamefont {M.}~\bibnamefont
		{Feygenson}}, \bibinfo {author} {\bibfnamefont {J.}~\bibnamefont {Carruth}},
	\bibinfo {author} {\bibfnamefont {R.}~\bibnamefont {Hoffmann}}, \ and\
	\bibinfo {author} {\bibfnamefont {K.~K.}\ \bibnamefont {Chipley}},\ }\href
{\doibase 10.1016/j.nimb.2012.05.037} {\bibfield  {journal} {\bibinfo
		{journal} {Nucl. Instrum. Meth. B}\ }\textbf {\bibinfo {volume} {287}},\
	\bibinfo {pages} {68 } (\bibinfo {year} {2012})}\BibitemShut {NoStop}%
\bibitem [{\citenamefont {Yang}\ \emph
	{et~al.}(2015{\natexlab{b}})\citenamefont {Yang}, \citenamefont {Juh\'{a}s},
	\citenamefont {Farrow},\ and\ \citenamefont {Billinge}}]{yang;arxiv15}%
\BibitemOpen
\bibfield  {author} {\bibinfo {author} {\bibfnamefont {X.}~\bibnamefont
		{Yang}}, \bibinfo {author} {\bibfnamefont {P.}~\bibnamefont {Juh\'{a}s}},
	\bibinfo {author} {\bibfnamefont {C.}~\bibnamefont {Farrow}}, \ and\ \bibinfo
	{author} {\bibfnamefont {S.~J.~L.}\ \bibnamefont {Billinge}},\ }\href
{http://arxiv.org/abs/1402.3163} {\bibfield  {journal} {\bibinfo  {journal}
		{arXiv}\ } (\bibinfo {year} {2015}{\natexlab{b}})},\ \bibinfo {note}
{1402.3163}\BibitemShut {NoStop}%
\bibitem [{\citenamefont {Farrow}\ \emph {et~al.}(2007)\citenamefont {Farrow},
	\citenamefont {Juh\'as}, \citenamefont {Liu}, \citenamefont {Bryndin},
	\citenamefont {{Bo\v zin}}, \citenamefont {Bloch}, \citenamefont {Proffen},\
	and\ \citenamefont {Billinge}}]{farro;jpcm07}%
\BibitemOpen
\bibfield  {author} {\bibinfo {author} {\bibfnamefont {C.~L.}\ \bibnamefont
		{Farrow}}, \bibinfo {author} {\bibfnamefont {P.}~\bibnamefont {Juh\'as}},
	\bibinfo {author} {\bibfnamefont {J.}~\bibnamefont {Liu}}, \bibinfo {author}
	{\bibfnamefont {D.}~\bibnamefont {Bryndin}}, \bibinfo {author} {\bibfnamefont
		{E.~S.}\ \bibnamefont {{Bo\v zin}}}, \bibinfo {author} {\bibfnamefont
		{J.}~\bibnamefont {Bloch}}, \bibinfo {author} {\bibfnamefont
		{T.}~\bibnamefont {Proffen}}, \ and\ \bibinfo {author} {\bibfnamefont
		{S.~J.~L.}\ \bibnamefont {Billinge}},\ }\href {\doibase
	10.1088/0953-8984/19/33/335219} {\bibfield  {journal} {\bibinfo  {journal}
		{J. Phys: Condens. Mat.}\ }\textbf {\bibinfo {volume} {19}},\ \bibinfo
	{pages} {335219} (\bibinfo {year} {2007})}\BibitemShut {NoStop}%
\bibitem [{\citenamefont {Juh\'{a}s}\ \emph {et~al.}(2015)\citenamefont
	{Juh\'{a}s}, \citenamefont {Farrow}, \citenamefont {Yang}, \citenamefont
	{Knox},\ and\ \citenamefont {Billinge}}]{juhas;aca15}%
\BibitemOpen
\bibfield  {author} {\bibinfo {author} {\bibfnamefont {P.}~\bibnamefont
		{Juh\'{a}s}}, \bibinfo {author} {\bibfnamefont {C.~L.}\ \bibnamefont
		{Farrow}}, \bibinfo {author} {\bibfnamefont {X.}~\bibnamefont {Yang}},
	\bibinfo {author} {\bibfnamefont {K.~R.}\ \bibnamefont {Knox}}, \ and\
	\bibinfo {author} {\bibfnamefont {S.~J.~L.}\ \bibnamefont {Billinge}},\
}\href {\doibase 10.1107/S2053273315014473} {\bibfield  {journal} {\bibinfo
	{journal} {Acta Crystallogr. A}\ }\textbf {\bibinfo {volume} {71}},\ \bibinfo
{pages} {562} (\bibinfo {year} {2015})}\BibitemShut {NoStop}%
\bibitem [{\citenamefont {Yang}\ \emph {et~al.}(2014)\citenamefont {Yang},
	\citenamefont {Juh\'{a}s},\ and\ \citenamefont {Billinge}}]{yang;jac14}%
\BibitemOpen
\bibfield  {author} {\bibinfo {author} {\bibfnamefont {X.}~\bibnamefont
		{Yang}}, \bibinfo {author} {\bibfnamefont {P.}~\bibnamefont {Juh\'{a}s}}, \
	and\ \bibinfo {author} {\bibfnamefont {S.~J.~L.}\ \bibnamefont {Billinge}},\
}\href {\doibase 10.1107/S1600576714010516} {\bibfield  {journal} {\bibinfo
	{journal} {J. Appl. Crystallogr.}\ }\textbf {\bibinfo {volume} {47}},\
\bibinfo {pages} {1273} (\bibinfo {year} {2014})}\BibitemShut {NoStop}%
\bibitem [{\citenamefont {Frandsen}\ \emph {et~al.}(2014)\citenamefont
	{Frandsen}, \citenamefont {Yang},\ and\ \citenamefont
	{Billinge}}]{frand;aca14}%
\BibitemOpen
\bibfield  {author} {\bibinfo {author} {\bibfnamefont {B.~A.}\ \bibnamefont
		{Frandsen}}, \bibinfo {author} {\bibfnamefont {X.}~\bibnamefont {Yang}}, \
	and\ \bibinfo {author} {\bibfnamefont {S.~J.~L.}\ \bibnamefont {Billinge}},\
}\href {\doibase 10.1107/S2053273313033081} {\bibfield  {journal} {\bibinfo
	{journal} {Acta Crystallogr. A}\ }\textbf {\bibinfo {volume} {70}},\ \bibinfo
{pages} {3} (\bibinfo {year} {2014})}\BibitemShut {NoStop}%
\bibitem [{\citenamefont {Frandsen}\ and\ \citenamefont
	{Billinge}(2015)}]{frand;aca15}%
\BibitemOpen
\bibfield  {author} {\bibinfo {author} {\bibfnamefont {B.~A.}\ \bibnamefont
		{Frandsen}}\ and\ \bibinfo {author} {\bibfnamefont {S.~J.~L.}\ \bibnamefont
		{Billinge}},\ }\href {\doibase 10.1107/S205327331500306X} {\bibfield
	{journal} {\bibinfo  {journal} {Acta Crystallogr. A}\ }\textbf {\bibinfo
		{volume} {71}},\ \bibinfo {pages} {325} (\bibinfo {year} {2015})}\BibitemShut
{NoStop}%
\bibitem [{\citenamefont {Frandsen}\ \emph {et~al.}(2016)\citenamefont
	{Frandsen}, \citenamefont {Brunelli}, \citenamefont {Page}, \citenamefont
	{Uemura}, \citenamefont {Staunton},\ and\ \citenamefont
	{Billinge}}]{frand;prl16}%
\BibitemOpen
\bibfield  {author} {\bibinfo {author} {\bibfnamefont {B.~A.}\ \bibnamefont
		{Frandsen}}, \bibinfo {author} {\bibfnamefont {M.}~\bibnamefont {Brunelli}},
	\bibinfo {author} {\bibfnamefont {K.}~\bibnamefont {Page}}, \bibinfo {author}
	{\bibfnamefont {Y.~J.}\ \bibnamefont {Uemura}}, \bibinfo {author}
	{\bibfnamefont {J.~B.}\ \bibnamefont {Staunton}}, \ and\ \bibinfo {author}
	{\bibfnamefont {S.~J.~L.}\ \bibnamefont {Billinge}},\ }\href {\doibase
	10.1103/PhysRevLett.116.197204} {\bibfield  {journal} {\bibinfo  {journal}
		{Phys. Rev. Lett.}\ }\textbf {\bibinfo {volume} {116}},\ \bibinfo {pages}
	{197204} (\bibinfo {year} {2016})}\BibitemShut {NoStop}%
\bibitem [{\citenamefont {Egami}\ and\ \citenamefont
	{Billinge}(2012)}]{egami;b;utbp12}%
\BibitemOpen
\bibfield  {author} {\bibinfo {author} {\bibfnamefont {T.}~\bibnamefont
		{Egami}}\ and\ \bibinfo {author} {\bibfnamefont {S.~J.~L.}\ \bibnamefont
		{Billinge}},\ }\href
{http://store.elsevier.com/product.jsp?lid=0\&iid=73\&sid=0\&isbn=9780080971414}
{\emph {\bibinfo {title} {Underneath the Bragg peaks: structural analysis of
			complex materials}}},\ \bibinfo {edition} {2nd}\ ed.\ (\bibinfo  {publisher}
{Elsevier},\ \bibinfo {address} {Amsterdam},\ \bibinfo {year}
{2012})\BibitemShut {NoStop}%
\bibitem [{\citenamefont {Frandsen}\ and\ \citenamefont
	{Billinge}(2016)}]{frand;unpub16}%
\BibitemOpen
\bibfield  {author} {\bibinfo {author} {\bibfnamefont {B.~A.}\ \bibnamefont
		{Frandsen}}\ and\ \bibinfo {author} {\bibfnamefont {S.~J.~L.}\ \bibnamefont
		{Billinge}},\ }\href@noop {} {\bibfield  {journal} {\bibinfo  {journal}
		{Unpublished}\ } (\bibinfo {year} {2016})}\BibitemShut {NoStop}%
\bibitem [{\citenamefont {Du~Tr\'{e}molet~de
		Lacheisserie}(1993)}]{dutre;b;mtaaom93}%
\BibitemOpen
\bibfield  {author} {\bibinfo {author} {\bibfnamefont {E.}~\bibnamefont
		{Du~Tr\'{e}molet~de Lacheisserie}},\ }\href@noop {} {\emph {\bibinfo {title}
		{Magnetostriction: theory and applications of magnetoelasticity}}}\ (\bibinfo
{publisher} {CRC Press},\ \bibinfo {address} {Boca Raton},\ \bibinfo {year}
{1993})\BibitemShut {NoStop}%
\bibitem [{\citenamefont {Sapega}\ \emph {et~al.}(2002)\citenamefont {Sapega},
	\citenamefont {Moreno}, \citenamefont {Ramsteiner}, \citenamefont
	{D\"aweritz},\ and\ \citenamefont {Ploog}}]{sapeg;prb02}%
\BibitemOpen
\bibfield  {author} {\bibinfo {author} {\bibfnamefont {V.~F.}\ \bibnamefont
		{Sapega}}, \bibinfo {author} {\bibfnamefont {M.}~\bibnamefont {Moreno}},
	\bibinfo {author} {\bibfnamefont {M.}~\bibnamefont {Ramsteiner}}, \bibinfo
	{author} {\bibfnamefont {L.}~\bibnamefont {D\"aweritz}}, \ and\ \bibinfo
	{author} {\bibfnamefont {K.}~\bibnamefont {Ploog}},\ }\href {\doibase
	10.1103/PhysRevB.66.075217} {\bibfield  {journal} {\bibinfo  {journal} {Phys.
			Rev. B}\ }\textbf {\bibinfo {volume} {66}},\ \bibinfo {pages} {075217}
	(\bibinfo {year} {2002})}\BibitemShut {NoStop}%
\bibitem [{\citenamefont {Luo}\ and\ \citenamefont {Martin}(2005)}]{luo;prb05}%
\BibitemOpen
\bibfield  {author} {\bibinfo {author} {\bibfnamefont {X.}~\bibnamefont
		{Luo}}\ and\ \bibinfo {author} {\bibfnamefont {R.~M.}\ \bibnamefont
		{Martin}},\ }\href {\doibase 10.1103/PhysRevB.72.035212} {\bibfield
	{journal} {\bibinfo  {journal} {Phys. Rev. B}\ }\textbf {\bibinfo {volume}
		{72}},\ \bibinfo {pages} {035212} (\bibinfo {year} {2005})}\BibitemShut
{NoStop}%
\bibitem [{\citenamefont {Jeong}\ \emph {et~al.}(2001)\citenamefont {Jeong},
	\citenamefont {Mohiuddin-Jacobs}, \citenamefont {Petkov}, \citenamefont
	{Billinge},\ and\ \citenamefont {Kycia}}]{jeong;prb01}%
\BibitemOpen
\bibfield  {author} {\bibinfo {author} {\bibfnamefont {I.}~\bibnamefont
		{Jeong}}, \bibinfo {author} {\bibfnamefont {F.}~\bibnamefont
		{Mohiuddin-Jacobs}}, \bibinfo {author} {\bibfnamefont {V.}~\bibnamefont
		{Petkov}}, \bibinfo {author} {\bibfnamefont {S.~J.~L.}\ \bibnamefont
		{Billinge}}, \ and\ \bibinfo {author} {\bibfnamefont {S.}~\bibnamefont
		{Kycia}},\ }\href {\doibase 10.1103/PhysRevB.63.205202} {\bibfield  {journal}
	{\bibinfo  {journal} {Phys. Rev. B}\ }\textbf {\bibinfo {volume} {63}},\
	\bibinfo {pages} {205202} (\bibinfo {year} {2001})}\BibitemShut {NoStop}%
\bibitem [{\citenamefont {Rana}\ and\ \citenamefont
	{Prasad}(2012)}]{rana;ijph12}%
\BibitemOpen
\bibfield  {author} {\bibinfo {author} {\bibfnamefont {N.~K.}\ \bibnamefont
		{Rana}}\ and\ \bibinfo {author} {\bibfnamefont {J.~N.}\ \bibnamefont
		{Prasad}},\ }\href {\doibase 10.1007/s12648-012-0095-1} {\bibfield  {journal}
	{\bibinfo  {journal} {Indian J. Phys.}\ }\textbf {\bibinfo {volume} {86}},\
	\bibinfo {pages} {601} (\bibinfo {year} {2012})}\BibitemShut {NoStop}%
\end{thebibliography}

%
	
\end{document}